\documentclass[lettersize, journal]{IEEEtran}

\usepackage[hidelinks]{hyperref}

\usepackage{float}
\usepackage{caption}
\usepackage{soul, color, xcolor}

\usepackage{cite}
\usepackage{amsmath,amssymb,amsfonts}
\expandafter\let\csname equation*\endcsname\relax
\expandafter\let\csname endequation*\endcsname\relax
\usepackage{amsthm}
\usepackage{graphicx}
\usepackage{textcomp}
\usepackage{svg}
\usepackage{subfigure}
\usepackage{algorithm}
\usepackage{mathtools}

\usepackage{algpseudocode}
\usepackage{subcaption}
\usepackage{subfigure} 

\newtheorem{theorem}{Theorem}

\newtheorem{lemma}{Lemma}

\newtheorem{corollary}{Corollary}

\newcounter{TempEqCnt}

\allowdisplaybreaks
\makeatletter
\def\ScaleIfNeeded{%
\ifdim\Gin@nat@ width>\linewidth \linewidth \else \Gin@nat@width
\fi } \makeatother

\graphicspath{{Fig/}}

\begin{document}

\title{Aerial Active STAR-RIS-Aided IoT NOMA Networks}

\author{Jingjing Zhao, 
Qian Xu, 
Xidong Mu, Yuanwei Liu,~\IEEEmembership{Fellow,~IEEE}, and Yanbo Zhu 
\thanks{J. Zhao, Q. Xu, and Y. Zhu are with the School of Electronics and Information Engineering, Beihang University, Beijing,
China (email:\{jingjingzhao, 15266317562, zhuyanbo\}@buaa.edu.cn).} 
\thanks{X. Mu is with the Centre for Wireless Innovation (CWI), Queen's University Belfast, Belfast, BT3 9DT, U.K. (e-mail: x.mu@qub.ac.uk).}
\thanks{Y. Liu is with the Department of Electrical and Electronic Engineering, the University of Hong Kong, Hong Kong, China (e-mail: yuanwei@hku.hk).}
}

\maketitle
\begin{abstract}

A novel framework of the unmanned aerial vehicle (UAV)-mounted active simultaneously transmitting and reflecting reconfigurable intelligent surface (STAR-RIS) communications with the non-orthogonal multiple access (NOMA) is proposed for Internet-of-Things (IoT) networks. In particular, an active STAR-RIS is deployed onboard to enhance the communication link between the base station (BS) and the IoT devices, and NOMA is utilized for supporting the multi-device connectivity. Based on the proposed framework, a system sum rate maximization problem is formulated for the joint optimization of the active STAR-RIS beamforming, the UAV trajectory design, and the power allocation. To solve the non-convex problem with highly-coupled variables, an alternating optimization (AO) algorithm is proposed to decouple the original problem into three subproblems. Specifically, for the active STAR-RIS beamforming, the amplification coefficient, the power-splitting ratio, and the phase shift are incorporated into a combined variable to simplify the optimization process. Afterwards, the penalty-based method is invoked for handling the non-convex rank-one constraint. For the UAV trajectory design and the power allocation subproblems, the successive convex optimization method is applied for iteratively approximating the local-optimal solution.
Numerical results demonstrate that: 1) the proposed algorithm achieves superior performance compared to the benchmarks in terms of the sum rate; and 2) the UAV-mounted active STAR-RIS can effectively enhance the channel gain from the BS to the IoT devices by the high-quality channel construction and the power compensation.
\end{abstract}


\section{Introduction}

The Internet of Things (IoT) network serves as a significant driver of social, economic, and technological advancements, catalyzing innovation across various sectors, including smart homes, industrial automation, healthcare, and smart cities~\cite{chen2014vision, frustaci2017evaluating, nguyen20216g}. These applications are instrumental in societal transformation and technological progress. However, the IoT devices are normally power-constrained, which limits the ability of data transmission over extended areas~\cite{zhao2023multiple}. To meet the extensive demands of IoT with limited resources, unmanned aerial vehicles (UAVs) offer an effective solution. Specifically, by acting as aerial base stations (BSs) or relays, UAVs can foster transmission flexibility, cost efficiency, and network coverage~\cite{lyu2016placement}.  

Nevertheless, the UAV-aided IoT networks face the following two main challenges. On the one hand, the signal quality should be guaranteed under the dynamic and uncertain communication environments \cite{zhou2024radar}. As a remedy, reconfigurable intelligent surface (RIS) is a promising technique to enhance the signal strength by adjusting the phases of electromagnetic waves \cite{zeng2020reconfigurable, zhao2022ris}. The traditional RIS can only support one-side reflection, thereby limiting its applicability in dynamic communication scenarios \cite{di2022communication}. With the capacity of enabling both transmission and reflection signals, the simultaneously transmitting and reflecting RIS (STAR-RIS) has been proposed to overcome the coverage limitation \cite{mu2021simultaneously, xu2021star,mu2024simultaneously}. Although STAR-RIS offers enhanced coverage and greater flexibility compared to the conventional RIS, both still experience the multiplicative path fading caused by the cascaded channels. To address this problem, the active STAR-RIS emerges these years \cite{mu2024simultaneously1,zhang2022active}. Compared to the passive counterpart, each active STAR-RIS element reflects/transmits the incident signals with the power compensation provided by the embedded amplifier~\cite{liu2022spectral}. As such, the active STAR-RIS can enhance both the signal strength and the transmission coverage. 
On the other hand, UAV-aided IoT networks face the challenge of massive connectivity, especially when a large number of devices require concurrent communications. Meanwhile, non-orthogonal multiple access (NOMA) has been envisioned to be a promising solution by superimposing signals at different power levels \cite{ding2015cooperative, ahsan2021resource,vu2021performance}. More specifically, by serving multiple users in the same time/frequency/code resource block, NOMA can significantly improve the spectrum efficiency, and thus accommodate the massive connectivity requirements of the IoT networks~\cite{duan2019resource}. 

\subsection{Related Works}

\subsubsection{Studies on UAV Communications} Extensive research works have been conducted on UAV-aided communications to enhance the transmission performance. The authors of \cite{motlagh2016low} discussed the key challenges of UAV deployment in IoT services and proposed an architecture designed to provide aerial value-added IoT services. Building on this, the paper \cite{mozaffari2017mobile} focused on the joint uplink power control, UAV mobility design, and device association in the time-varying IoT networks. Further advancing this field, the authors of \cite{chen2021joint} investigated a deployment scheme with Quality of Service (QoS) guarantees, and strategically placed multiple UAVs to enhance the average data rate. 
The outage probability was derived by the authors of \cite{rupasinghe2018non} for the UAV-aided IoT networks employing the NOMA technique. In \cite{wu2018capacity}, the authors considered two special cases to characterize the capacity region of a UAV-enabled broadcast channel within given flight time. Furthermore, the authors of \cite{duan2019resource} studied both of the UAV flying height and sensor node transmit power optimization problems to enhance system capacity in the multi-UAV scenarios. The authors of \cite{zhao2020security} proposed a secure transmission scheme for the UAV-NOMA networks by optimizing the UAV position and the beamforming. In \cite{masaracchia2020energy}, the authors introduced a time-sharing NOMA scheme for the UAV-aided system to improve downlink fairness and spectral efficiency. A general NOMA-enable data collection protocol was proposed in a wireless sensor network by the authors of \cite{chen2020uav}, where a sum rate optimization problem was formulated. 
The authors of \cite{katwe2021dynamic} developed a two-stage dynamic user clustering strategy to enhance throughput in the UAV-assisted NOMA system.
Moreover, Generative AI Agents, leveraging large language models (LLMs) and related technologies, can further enhance UAV-based IoT networks by providing intelligent solutions to specific problems~\cite{10531073}.


\subsubsection{Studies on RIS-Aided Communications} To reap the benefits of RISs in terms of enhanced channel quality, the integration of RISs into UAV communications has attracted some research contributions recently. A sum-rate optimization problem was characterized by the authors of \cite{wei2020sum} for a RIS-aided UAV network employing orthogonal frequency division multiple access (OFDMA). The authors of \cite{li2020reconfigurable} proposed a novel approach to optimize the average achievable rate of the RIS-aided UAV communication systems by employing iterative design for both of the UAV trajectory and the RIS passive beamforming. Assuming that the eavesdropper's channel state information is perfectly known, a secure communication framework was developed in \cite{li2021reconfigurable}  
aiming at maximizing the average secrecy rate. Additionally, the high mobility of UAV and the tunable capability of RIS were utilized by the authors of \cite{pang2021irs} to defend against eavesdroppers in complex urban scenarios. The authors of \cite{mu2021intelligent} focused on maximizing the sum rate in RIS-aided multi-UAV NOMA networks, and demonstrated that the proposed algorithm outperformed other benchmark schemes. 
The authors of \cite{10032267} proposed a proximal policy optimization (PPO)-based approach for energy efficiency maximization in RIS-assisted simultaneous wireless information and power transfer networks with rate splitting multiple access.
The authors of \cite{zhao2022simultaneously} studied a UAV communication system aided by a STAR-RIS, and demonstrated that the STAR-RIS provided better performance than the conventional RIS. In \cite{guo2023secure}, the authors investigated the secrecy energy efficiency maximization problem in the mobile scenarios, and employed a UAV-mounted STAR-RIS to counteract eavesdroppers. However, STAR-RIS typically operates with passive elements, which may limit its performance. 
To address this limitation, the active STAR-RIS integrates components to amplify incident signals. In \cite{xu2023active}, a hardware model was studied, deriving amplitude gain for independent and coupled reflection/transmission shifts. The authors of \cite{xu2023active} and \cite{10264149} also explored energy efficiency and cost-effectiveness, providing guidelines to balance performance and resource constraints. In \cite{10264149}, fractional power control was proposed in an active STAR-RIS aided multiple-input multiple-output (MIMO) system to improve spectral and energy efficiency.


\subsection{Motivations and Contributions}

 Although previous works have established a sturdy foundation for UAV and RIS-enabled communications, the exploration of adopting active STAR-RIS in UAV-aided IoT networks remains largely unexplored. As unveiled by the recent works \cite{xu2023active, 10227341, 10264149}, the active STAR-RIS can overcome the ``double-fading" effects and enhance the beamforming gain. According to our best knowledge, there has been no prior research exploring the potential performance gain of UAV-mounted active STAR-RIS in IoT networks employing NOMA, for which the primary challenges can be described as: 1) The non-convex optimization problem is challenging to solve because of the highly-intertwined optimization variables, including the UAV trajectory, the active STAR-RIS amplification gain and the reflection/transmission coefficients, as well as the power allocation coefficients; 2) The integration of NOMA adds additional complexity in the form of condition-based decoding order design, leading to intricate interplay between the subproblems. Thus, the development of an efficient algorithm is crucial to leverage the potential of deploying the active STAR-RIS in the UAV-aided IoT NOMA networks. 

 The primary contributions of this paper are detailed as follows:
\begin{itemize}
    \item We investigate a transmission framework for the UAV-IoT networks, in which an active STAR-RIS is deployed on the UAV to enhance the channel quality between the UAV and the IoT devices, and NOMA is utilized at the BS to serve multiple ground IoT devices. Based on this framework, we formulate a sum rate maximization problem, aiming for the joint optimization of the UAV trajectory, the active STAR-RIS amplification gain and reflection/transmission coefficient, as well as the power allocation.

    \item We develop an alternating optimization (AO) algorithm, where the original problem is decomposed into three subproblems that are solved alternatingly. For the active STAR-RIS beamforming subproblem, we construct the combined variables which incorporate the amplification coefficient, the power-splitting ratio and the phase shift, so as to simplify the optimization process. Afterwards, the penalty-based method is invoked for handling the non-convex constraint. For both the UAV trajectory design and the power allocation subproblems, we effectively solve them by utilizing the successive convex approximation (SCA) technique while the other optimization variables are fixed.
    
    
    \item  Our numerical results indicate that the proposed algorithm achieves superior sum rate performance compared to the benchmark schemes for the UAV-aided active IoT network. It is demonstrated that the UAV-mounted active STAR-RIS can significantly boost the channel quality between the BS and the IoT devices by reconstructing high-quality channels and providing effective power compensation. Moreover, the NOMA gain over OMA is distinctly revealed, thanks to the enlarged channel differences with the flexible UAV trajectory and the channel reconstruction through the active STAR-RIS. 

\end{itemize}

\subsection{Organization and Notation}
The remaining structure of this paper is arranged as follows. In Section~II, the system model is first introduced, which is followed by the sum rate maximization problem formulation. In Section III, an AO-based iterative algorithm is developed to address the joint beamforming, trajectory design, and power allocation problem. Section IV presents the simulation results, and Section V concludes the paper with final remarks. 

\textit{Notation}: Scalars are represented by italic letters, vectors by bold lower-case letters, and matrices by bold upper-case letters. $\mathbb{C}^{N\times 1}$ denotes the space of $N\times 1$ complex-valued vectors. For a vector $\mathbf{a}$, $\mathbf{a}^*$denotes its conjugate, $\mathbf{a}^H$ denotes its (Hermitian) conjugate transpose, and $||\mathbf{a}||$ denotes its the Euclidean norm. $\text{diag}(\mathbf{a})$ denotes a diagonal matrix, in which the main diagonal elements are the vector $\mathbf{a}$'s elements. $\mathcal{C}\mathcal{N}\left(\mu,\sigma^2\right)$ denotes the distribution of a circularly symmetric complex Gaussian (CSCG) random variable with a mean of $\mu$ and a variance of $\sigma^2$. $\mathbb{H}^N$ denotes the set of all $N$-dimensional complex Hermitian matrices. For a matrix $\mathbf{X}$, $\text{Rank}(\mathbf{X})$ denotes its rank, and $\text{Tr}(\mathbf{X})$ denotes its trace. $\text{Diag}(\mathbf{X})$ denotes a vector formed by the main diagonal elements of matrix $\mathbf{X}$. $\mathbf{X} \succeq 0$ signifies that $\mathbf{X}$ is positive semidefinite. The norms $||\mathbf{X}||_*$, $||\mathbf{X}||_2$, and $||\mathbf{X}||_F$ represent the nuclear, spectral, and Frobenius norms of the matrix, respectively. $\mathbf{1}_{M\times1}$ denotes an $M\times1$ vector with all elements equal to 1.

\section{System Model and Problem Formulation}

In this section, we first introduce the system model of the UAV-aided IoT network incorporating the active STAR-RIS. Then, the joint optimization problem of the UAV trajectory design, the active STAR-RIS beamforming, and the power allocation is formulated.



\subsection{System Description}
\begin{figure}
    \centering
    \includegraphics[scale=0.60]{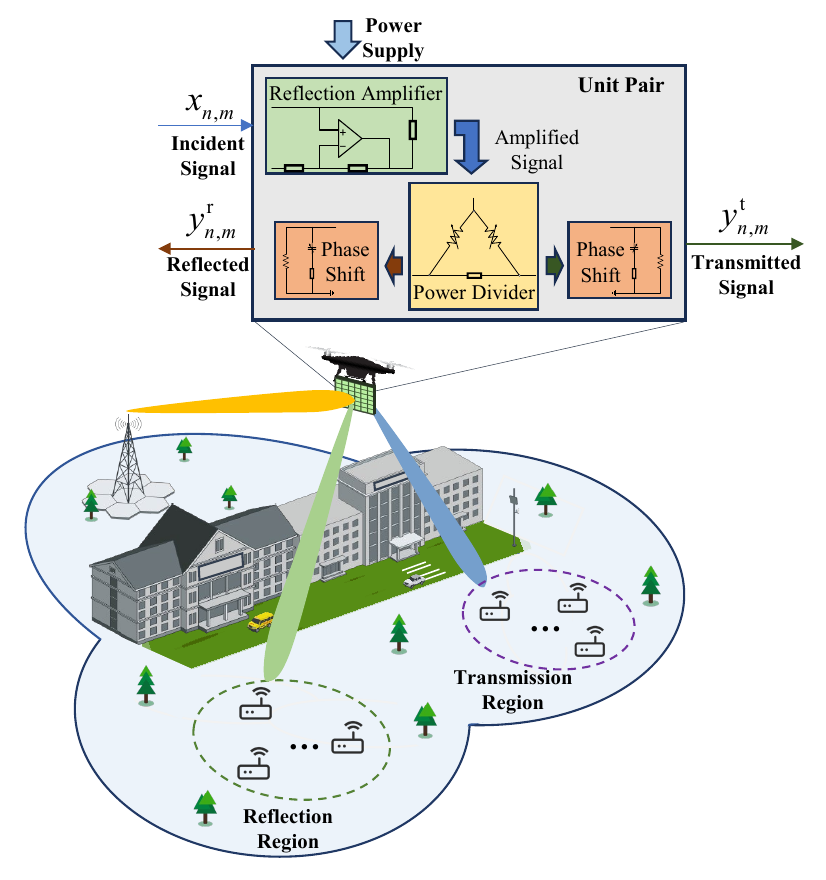}
    \caption{System model of the UAV-mounted active STAR-RIS-aided IoT networks.}
    \label{fig: system model}
\end{figure}

Consider a UAV-aided IoT network, consisting of one UAV-mounted active STAR-RIS, one BS, and $K$ ground IoT devices. As shown in Fig. \ref{fig: system model}, due to the obstacles, the LoS/non line-of-sight (NLoS) communication links between the BS and the ground IoT devices are blocked. We assume that the perfect CSI of all channels is assumed to be available at the BS to study the maximum performance. In practice, approaches as presented in \cite{wu2021channel, shtaiwi2021channel} can be deployed in our work for channel estimation with acceptable complexity and overhead. The BS and the IoT devices are all equipped with a signal antenna. Assume that the number of IoT devices located in the reflection and transmission region of the active STAR-RIS is $R$ and $T$, respectively. The overall set of IoT devices is represented by $\mathcal{K}=\{1, 2,..., K\}$, where $K=R+T$. $\mathcal{K}_{\text{r}}$ and $\mathcal{K}_{\text{t}}$ denote the set of IoT devices in the reflection region and transmission region, respectively, which satisfy $\mathcal{K}_{\text{r}}\cap\mathcal{K}_{\text{t}}=\varnothing$ and $\mathcal{K}_{\text{r}}\cup\mathcal{K}_{\text{t}}=\mathcal{K}$. The active STAR-RIS consists of $M=M_{\text{v}}\times M_{\text{h}}$ elements, and each element consists of a reflection amplifier, a power divider, and two-phase shifts to direct the incident signal to the desired direction. Denote the amplification gain matrix as $\mathbf{A}_n\triangleq\text{diag}({\boldsymbol{\alpha}_n})$ with $\boldsymbol{\alpha}_n\triangleq[\sqrt{\alpha_{n, 1}},...,\sqrt{\alpha_{n, M}}]^{T}$, where $\sqrt{\alpha_{n,m}}$ is the amplification coefficient of each element. Furthermore, the reflection and transmission amplitude coefficients are represented by $\mathbf{E}_{n}^{\text{r}}\triangleq\text{diag}\left([\varsigma_{n,1},..,\varsigma_{n,M}]\right)$, and $\mathbf{E}_{n}^{\text{t}}\triangleq\text{diag}\left(\left[\sqrt{1-\varsigma_{n,1}^2},..,\sqrt{1-\varsigma_{n,M}^2}\right]\right)$, respectively, where $\varsigma_{n,m}\in[0,1]$ is the reflection amplitude. Each STAR-RIS element maintains a balance between reflection and transmission, as dictated by the law of energy conservation. Specifically, the relationship $\varsigma_{n,m}^2 + (\sqrt{1-\varsigma_{n,m}^2}^2) = 1$ ensures that the total energy processed by each element is conserved \cite{xu2021star}.
We further denote the reflection and transmission phase-shift matrices as
$\boldsymbol{\Phi}_{n}^{\text{r}}\triangleq{\text{diag}}\left(\left[\phi_{n,1}^{\text{r}},...,\phi_{n,M}^{\text{r}}\right]\right)$ and $\boldsymbol{\Phi}_{n}^{\text{t}}\triangleq{\text{diag}}\left(\left[\phi_{n,1}^{\text{t}},...,\phi_{n,M}^{\text{t}}\right]\right)$, respectively. Here, $\phi_{n,m}^{\text{r}}=e^{j\theta_{n,m}^{\text{r}}}$ and $\phi_{n,m}^{\text{t}}=e^{j\theta_{n,m}^{\text{t}}}$ are the phase shifts introduced by the $m$-th active STAR-RIS element to the reflected and transmitted signals, respectively.

Note that the active STAR-RIS amplification circuit must operate within its linear range, ensuring that the output power increases in direct proportion to the input power \cite{you2021wireless, ma2023optimization}. As such, we incorporate the per-element power constraint during time slot $n$, which is given by
\begin{equation}
\label{eq: element-wise power constraint of system model}
  P_\text{B}^\text{max}\alpha_{n,m}|h_{n,m}^{\text{bs}}|^{2}+\alpha_{n,m}\sigma_{\text{v}}^{2}\leq p_{n,m}^{\text{max}}, \forall{m},
\end{equation}
where $P_\text{B}^\text{max}$ denotes the total power constraint of the BS, $\mathbf{v}_{n} \sim \mathcal{C}\mathcal{N}\left(0,\sigma_{\text{v}}^2\mathbf{1}_{M\times1}\right)$ denotes the noise vector over all the active STAR-RIS elements, and $p_{n,m}^{\text{max}}$ denotes the power constraint of the $m$-th active STAR-RIS element. Additionally, considering the amplification gain matrix $\mathbf{A}_n$ of the active STAR-RIS, the overall power constraint can be given by 
\begin{equation}
\label{eq: total power constraint of system model}
  P_\text{B}^\text{max}||\mathbf{A}_n\mathbf{h}_{n}^{\text{bs}}||_{2}^{2}+\sigma_{\text{v}}^{2} ||\mathbf{A}_n||_{\text{F}}^{2}\leq P_{n}^{\text{max}},
\end{equation}
where $P_{n}^{\text{max}}$ is the total power limit for the active STAR-RIS, with the constraint $P_{n}^{\text{max}} \leq \sum_{m=1}^M p_{n,m}^{\text{max}}$ imposed due to the thermal constraints of the circuit.

Assuming a three-dimensional (3D) Cartesian coordinate system, the location of the BS and the $k$-th IoT device are denoted by $\mathbf{q}_{\text{B}}=\left\{{x_{\text{B}},y_{\text{B}},z_{\text{B}}}\right\}^{T}$, $\mathbf{q}_k=\left\{x_k,y_k,0\right\}^{T}$, $k\in \mathcal{K}$, respectively. For simplicity of analysis, the UAV's flight time $T$ is divided into $N$ time slots of equal length, where each slot has a duration of $\delta=\frac{T}{N}$. During time slot $n$, the position of the UAV is denoted by $\mathbf{q}_{\text{u}}(n)=\left\{{x_{\text{u}}(n),y_{\text{u}}(n),z_{\text{u}}(n)}\right\}^{T}$. The position of the active STAR-RIS center is represented by $\mathbf{q}_{\text{R}}(n)=\left\{x_{\text{R}}(n),y_{\text{R}}(n),z_{\text{R}}(n)\right\}^{T}$. The UAV flies at a fixed altitude of $z_{\text{u}}(n)=H$ and has a maximum speed of $V_{\text{max}}$. Between two adjacent time slots, the UAV's position must satisfy the following constraints:

\begin{subequations}
\begin{equation}
\label{eq: A_trajectory}
    ||\mathbf{q}_{\text{u}}(n+1)-\mathbf{q}_{\text{u}}(n)||^{2}\leq D^{2}, n=1,2,...,N-1,
\end{equation}
\begin{equation}
\label{eq: B_trajectory}
    ||\mathbf{q}_{\text{u}}(1)-\mathbf{q}_{\text{u}}^{0}||^{2}\leq D^{2},
\end{equation}
\begin{equation}
\label{eq: C_trajectory}
    ||\mathbf{q}_{\text{u}}^{\text{F}}-\mathbf{q}_{\text{u}}(n)||^{2}\leq D^{2},
\end{equation}
\end{subequations}
where $q_{\text{u}}^{0}$ and $q_{\text{u}}^{\text{F}}$ denote the UAV's start and final positions, respectively, and $D=V_{\text{m}}\delta$ is the maximum distance which the UAV can travel in a single time slot.

\subsection{Channel and Signal Model}
Assume that the active STAR-RIS operates in the energy splitting (ES) mode, where the energy of the incident signal on each element is divided between the transmitted and reflected signals. Then, the transmitted and reflected signals can be modeled individually as
\begin{equation}
\label{eq: reflected signal of receiver}
   y_{n,m}^{\text{r}}=\phi_{n,m}^{\text{r}}\varsigma_{n,m}\sqrt{\alpha_{n,m}}\left(x_{n,m}+v_{m}\right),
\end{equation}
\begin{equation}
\label{eq: transmit signal of receiver}
   y_{n,m}^{\text{t}}=\phi_{n,m}^{\text{t}}\sqrt{1-\varsigma_{n,m}^2}\sqrt{\alpha_{n,m}}\left(x_{n,m}+v_{m}\right),
\end{equation}
where $x_{n,m}$ is the signal incident on the $m$-th element of the active STAR-RIS, and $v_{m}$ is the noise that the active STAR-RIS introduces.

The signal received by the $k$-th IoT device can be expressed as
\begin{equation}
\label{eq: reflected signal IoT device i}
    \begin{split}
   y_{n,k}^p= &\left(\mathbf{h}_{n,k}^{p}\right)^{H}\boldsymbol{\Phi}_{n}^{p}\mathbf{E}_{n}^{p}\mathbf{A}_n\mathbf{h}_{n}^{\text{bs}}x_{n} \\
   &+\left(\mathbf{h}_{n,k}^{p}\right)^{H}\boldsymbol{\Phi}_{n}^{p}\mathbf{E}_{n}^{p}\mathbf{A}_n\mathbf{v}_{n}+n_{n, k}, \forall {p}\in\{\text{r},\text{t}\},
    \end{split}
\end{equation}
where $x_{n}$ is the transmit signal at the BS,  $\mathbf{h}_{n}^{\text{bs}}\in \mathbb{C}^{M\times1}$, $\mathbf{h}_{n,k}^{\text{r}}\in \mathbb{C}^{M\times1}$, $\mathbf{h}_{n,k}^{\text{t}}\in \mathbb{C}^{M\times1}$ denote the channel links between the BS and the active STAR-RIS, that between the active STAR-RIS and the device in the reflection region, and that between the active STAR-RIS and the device in the transmission region, respectively. Moreover, $n_{n, k}\sim \mathcal{C}\mathcal{N}\left(0,\sigma_{n, k}^2\right)$ represents the noise at the IoT device.

As the UAV flies at a high-altitude, it is highly likely that the LoS channel components exist between the BS and the active STAR-RIS, as well as between the active STRA-RIS and the IoT devices. Thus, the channels $\mathbf{h}_{n}^{\text{bs}}$ and $\mathbf{h}_{n,k}^{p}$ can be modeled using the Rician fading distribution, which can be formulated as 
\begin{equation}
\label{eq: BS to STAR-RIS channel}
    \mathbf{h}_{n}^{\text{bs}}= \sqrt{\frac{\rho_0}{d_{\text{bs}}(n)^{\tau_0}}} \left(\sqrt{\frac{\beta_{\text{bs}}}{1+\beta_{\text{bs}}}}\mathbf{h}_{n}^{\text{bs-LoS}}+\sqrt{\frac{1}{1+\beta_{\text{bs}}}}\mathbf{h}_{n}^{\text{bs-NLoS}}\right),
\end{equation}
\begin{equation}
\label{eq: STAR-RIS to reflected signal IoT device r channel}
    \mathbf{h}_{n,k}^{p}= \sqrt{\frac{\rho_0}{d_{\text{s}k}(n)^{\tau_0}}} \left(\sqrt{\frac{\beta_{\text{s}}}{1+\beta_{\text{s}}}}\mathbf{h}_{n,k}^{p-\text{LoS}}+\sqrt{\frac{1}{1+\beta_{\text{s}}}}\mathbf{h}_{n,k}^{p-\text{NLoS}}\right), 
\end{equation}
respectively, where ${p}\in\{\text{r},\text{t}\}$, $\rho_0$ denotes the path loss at the reference distance of 1 meter, $d_{\text{bs}}(n)=||\mathbf{q}_{\text{B}}-\mathbf{q}_{\text{u}}(n)||$ denotes the distance between the BS and the active STAR-RIS during time slot $n$, $d_{\text{s}k}(n)=||\mathbf{q}_{\text{u}}(n)-\mathbf{q}_k||$ denotes the distance between the active STAR-RIS and the IoR devices during time slot $n$, $\tau_0$ denotes the path loss exponents, $\beta_{\text{bs}}$ and $\beta_{\text{s}}$ denote the Rician factor of the BS - active STAR-RIS link, and that of the active STAR-RIS - IoT devices links, respectively. $\mathbf{h}_{n}^{\text{bs-LoS}}=\boldsymbol{\alpha}\left(\omega^{\text{bs}},\psi^{\text{bs}}\right)$ and $\mathbf{h}_{n,k}^{p-\text{LoS}}=\boldsymbol{\alpha}\left(\omega^{p},\psi^{p}\right)$ are the deterministic LoS components, where $\boldsymbol{\alpha}(\omega,\psi)$ represents the array response vector (ARV). $\mathbf{h}_{n}^{\text{bs-NLoS}}$ and $\mathbf{h}_{n,i}^{p-\text{NLoS}}$ denote the NLoS components. For $\boldsymbol{\alpha}(\omega,\psi)$, the expression can be given by
\begin{equation}
\label{eq: array response vector}
    \begin{split}
       \boldsymbol{\alpha}(\omega,\psi)=&\big[1,\cdots,e^{-j\frac{2\pi d_{I}}{\lambda}(m_v -1)\sin{\omega}\cos{\psi}},\cdots, \\
       & e^{-j\frac{2\pi d_{I}}{\lambda}(M_v -1)\sin{\omega}\cos{\psi}}\big]^T\\
       &\otimes\big[1,\cdots,e^{-j\frac{2\pi d_{I}}{\lambda}(m_h -1)\sin{\omega}\sin{\psi}},\cdots,\\
       & e^{-j\frac{2\pi d_{I}}{\lambda}(M_h -1)\sin{\omega}\sin{\psi}}\big]^T,
    \end{split}
\end{equation}
where $\omega\in\left[-\frac{\pi}{2},\frac{\pi}{2}\right]$ represents the elevation angle-of-arrival (AoA)/angle-of-departure (AoD), and $\psi\in[0,2\pi]$ represents the azimuth AoA/AoD.
As such, the overall channel from the BS to the $k$-th IoT device can be expressed as 
\begin{equation}
\label{eq: reflected channel gain before square}
   g_{n,k}=\left(\mathbf{h}_{n,k}^{p}\right)^{H}\boldsymbol{\Phi}_{n}^{p}\mathbf{E}_{n}^{p}\mathbf{A}_n\mathbf{h}_{n}^{\text{bs}}, \forall {p}\in\{\text{r},\text{t}\}.
\end{equation}
According to the NOMA principle, each IoT device utilizes successive interference cancellation (SIC) to eliminate intra-cell interference. Let $\mu_n(k) \in \left\{1,...,K\right\}$ represent the decoding order of IoT device $k$ at time slot $n$. For any two IoT devices $k$ and $l$ where $\mu_n(k) > \mu_n(l)$, IoT device $k$ must decode the signal of device $l$ prior to decoding its own. The decoding sequence is followed under the condition that the combined channel power gains satisfy $|g_{n,k}|^{2}\geq|g_{n,l}|^{2}$. Furthermore, the condition $p_{n,l} \geq p_{n,k} \geq 0$ is applied to ensure fairness among IoT devices, where $p_{n,k}$ represents the transmission power allocated by the BS to the $k$-th IoT device. Note that since the position of the UAV influences the combined channel power gain, any of the $K!$ possible decoding order combinations can occur. As such, the set of all possible decoding sequences is represented by $\mathcal{D}$, where $|\mathcal{D}|=K!$. To simplify the complexity for decoding the transmission framework, we will present an efficient decoding order design scheme in Section III-D.

With given decoding order, the signal-to-noise-plus-interference ratio (SINR) of the $k$-th IoT device can be given by 
\begin{equation}
\label{eq: SINR of reflected IoT device i}
   \gamma_{n,k}^{p}=\frac{|g_{n,k}|^{2}p_{n,k}}{|g_{n,k}|^{2}
   \sum\limits_{\mu(i)>\mu(k)} p_{n,i} +\sigma_{\text{v}}^2||\mathbf{A}_n\mathbf{E}_{n}^{p}\mathbf{h}_{n,k}^{p}||_2^2+\sigma_{n, k}^2},
\end{equation}
where $||(\boldsymbol{\Phi}_n^{p})^*\mathbf{A}_n\mathbf{E}_n^{p}\mathbf{h}_{n,k}^{p}||_2^2=||\mathbf{A}_n\mathbf{E}_{n}^{p}\mathbf{h}_{n,k}^{p}||_2^2$ with $p\in\{\text{r}, \text{t}\}$.
Thus, in the $n$-th time slot, the data rate of the $k$-th IoT device can be expressed as  
\begin{equation}
\label{eq: data rate of reflected IoT device i}
   R_{n,k}^{p}=\log_{2}\left(1+\gamma_{n,k}^{p}\right), \forall {p}\in\{\text{r}, \text{t}\}.
\end{equation}

\subsection{Problem Formulation}
\label{sec: Problem Formulation}
Denote $\mathbf{Q}=\left\{\mathbf{q}_{\text{u}}(n), \forall n\right\}$ as the UAV trajectory matrix, $\mathbf{A}=\left\{\mathbf{A}_n,\forall n\right\}$ as the amplification gain matrix, $\boldsymbol{\Phi}^{\text{r}}=\left\{\boldsymbol{\Phi}_{n}^{\text{r}},\forall n\right\}$ and $\boldsymbol{\Phi}^{\text{t}}=\left\{\boldsymbol{\Phi}_{n}^{\text{t}},\forall n\right\}$ as the reflected and transmitted phase shifts matrix, respectively, $\mathbf{E}^{\text{r}}=\left\{\mathbf{E}^{\text{r}}_n, \forall n\right\}$ and $\mathbf{E}^{\text{t}}=\left\{\mathbf{E}^{\text{t}}_n, \forall n\right\}$ as the reflection and transmission amplitude coefficients at the active STAR-RIS, respectively, and $\mathbf{P}=\left\{p_{nk}, \forall n,k\right\}$ as the power allocation coefficients at the BS. As a result, the corresponding optimization problem is reformulated as

\begin{subequations}
\label{eq: optimization_problem}
\begin{equation}
\label{eq: MSE_objective}
    \max_{\left\{\mathbf{Q}, \mathbf{A}, \boldsymbol{\Phi}^{\text{r}}, \boldsymbol{\Phi}^{\text{t}}, \mathbf{E}^{\text{r}}, \mathbf{E}^{\text{r}}, \mathbf{P} \right\}} \sum_{n=1}^{N} \sum_{k\in\mathcal{K}} R_{n,k}^{p} 
\end{equation}
\begin{equation}
\label{eq: A_element-wise power constraint}
  {\rm{s.t.}} \ P_\text{B}^\text{max}\alpha_{n,m}|h_{n,m}^{\text{bs}}|^{2}+\alpha_{n,m}\sigma_{\text{v}}^{2}\leq p_{n,m}^{\text{max}}, \forall{m}, n,
\end{equation}
\begin{equation}
\label{eq: B_total power constraint}
  P_\text{B}^\text{max}||\mathbf{A}_n\mathbf{h}_{n}^{\text{bs}}||_{2}^{2}+\sigma_{\text{v}}^{2} ||\mathbf{A}_n||_{\text{F}}^{2}\leq P_{n}^{\text{max}},\forall n,
\end{equation}
\begin{equation}
\label{eq: reflected & transmitted coefficient constraint}
    \varsigma_{n,m}\in [0,1], \forall m, n,
\end{equation}
\begin{equation}
\label{eq: amplified coefficient constraint}
    \boldsymbol{\alpha}_n\geq 1, \forall m, n,
\end{equation}
\begin{equation}
\label{eq: shift constraints}
    |\phi_{n,m}^{\text{r}}|=1, |\phi_{n,m}^{\text{t}}|=1, \forall m, n,
\end{equation}
\begin{equation}
\label{eq: decoding strategy of channel gain}
    |g_{n,k}|^{2}\geq|g_{n,l}|^{2}, \text{if}\ \mu_n(k) > \mu_n(l),
\end{equation}
\begin{equation}
\label{eq: power allocation by decoding order}
    0\leq p_{n,k}\leq p_{n,l}, \text{if}\ \mu_n(k) > \mu_n(l),
\end{equation}
\begin{equation}
\label{eq: total power allocation}
    \sum_{k=1}^K p_{n,k}\leq P_\text{B}^\text{max}, \forall n,
\end{equation}
\begin{equation}
\label{eq: trajectory}
    \eqref{eq: A_trajectory} \sim \eqref{eq: C_trajectory},
\end{equation}
\end{subequations}
\eqref{eq: A_element-wise power constraint} and \eqref{eq: B_total power constraint} are the maximum power constraint of each element and the total power constraint of the active STAR-RIS, respectively. \eqref{eq: reflected & transmitted coefficient constraint}-\eqref{eq: shift constraints} are the power-split, amplitude, and phase shift constraints of the active STAR-RIS, respectively. Moreover, \eqref{eq: decoding strategy of channel gain} and \eqref{eq: power allocation by decoding order} are the SIC decoding order constraints. \eqref{eq: total power allocation} is the maximum total transmit power constraint at the BS, and \eqref{eq: trajectory} is the UAV trajectory constraint. 

Problem \eqref{eq: optimization_problem} is intractable to solve given that the optimization variables are highly coupled and the problem is neither convex or concave. It is worth noting that, compared to the sum rate maximization problem for the conventional passive STAR-RIS aided communications, the amplification coefficient matrix $\mathbf{A}$ of the active STAR-RIS needs to be jointly optimized with the other optimization variables, which makes the problem more complex. Moreover, due to the thermal noise introduced by the active elements, the beamforming requires the consideration of its effect on the noise at the active STAR-RIS, which further complicates the optimization problem.

\setcounter{TempEqCnt}{\value{equation}} 
\setcounter{equation}{17} 
\begin{figure*}[ht]
    \begin{equation}
    \label{eq: first-order Taylor of Beamforming objective function}
        \log_2\left(1+\frac{1}{c_{n,k}d_{n,k}}\right)\geq\log_2\left(1+\frac{1}{c_{n,k}^{(\tau_1)}d_{n,k}^{(\tau_1)}}\right)-\frac{\log_2e(c_{n,k}-c_{n,k}^{(\tau_1)})}{c_{n,k}^{(\tau_1)}\left(1+c_{n,k}^{(\tau_1)}d_{n,k}^{(\tau_1)}\right)}-\frac{\log_2e(d_{n,k}-d_{n,k}^{(\tau_1)})}{d_{n,k}^{(\tau_1)}\left(1+c_{n,k}^{(\tau_1)}d_{n,k}^{(\tau_1)}\right)}\triangleq \tilde{R}_{n,k}^{p},
    \end{equation}
    \hrulefill
\end{figure*}
\setcounter{equation}{\value{TempEqCnt}} 

\section{AO-BASED ALTERNATING OPTIMIZATION ALGORITHM}
In this section, we propose an AO algorithm, that decomposes the original optimization problem into the active STAR-RIS beamforming, the UAV trajectory design, and the power allocation subproblems. Each group of variables is iteratively optimized while keeping the others fixed, resulting in an iterative optimization process. 

To enable the implementation of the AO-based algorithm, we first partition the optimization variables of problem \eqref{eq: optimization_problem} into three groups: $\{\mathbf{A}, \boldsymbol{\Phi}^{\text{r}}, \boldsymbol{\Phi}^{\text{t}}, \mathbf{E}^{\text{r}}, \mathbf{E}^{\text{t}}\}$, $\{\mathbf{Q}\}$, and $\{\mathbf{P}\}$. For the optimization of $\{\mathbf{A}, \boldsymbol{\Phi}^{\text{r}}, \boldsymbol{\Phi}^{\text{t}}, \mathbf{E}^{\text{r}}, \mathbf{E}^{\text{t}}\}$, we construct the new optimization variables $\mathbf{U}_n^{\text{r}}$ and $\mathbf{U}_n^{\text{t}}$ by combining the amplification coefficient, the power-slitting ratio and the phase-shift into one variable, so as to reduce the solving complexity. The penalty-based approach and the SCA method are applied for addressing each subproblem iteratively. 

\subsection{Active STAR-RIS Amplification Gain and Reflection/Transmission Coefficient Optimization}
\label{sec:coordinate-transformation}
We start by reformulating problem \eqref{eq: optimization_problem} into a more manageable form. To simplify the design process, we introduce the reflection/transmission coefficient vectors defined as 
$\mathbf{u}_n^{\text{r}}=\left[\sqrt{\alpha_{n,1}}\varsigma_{n,1}e^{j\theta_{n,1}^{\text{r}}},\cdots,\sqrt{\alpha_{n,M}}\varsigma_{n,M}e^{j\theta_{n,M}^{\text{r}}}\right]$ and $\mathbf{u}_n^{\text{t}}=\left[\sqrt{\alpha_{n,1}}\sqrt{1-\varsigma_{n,1}^2}e^{j\theta_{n,1}^{\text{t}}},\cdots,\sqrt{\alpha_{n,M}}\sqrt{1-\varsigma_{n,M}^2}e^{j\theta_{n,M}^{\text{t}}}\right]$, which leads to $|g_{n,k}|^2=\left|(\mathbf{u}_n^{p})^H\mathbf{h}_{n,k}^{\text{bg}-{p}}\right|^2$, $\forall {p}\in \{{\text{r}},{\text{t}}\}$, where $\mathbf{h}_{n,k}^{\text{bg}-{p}}=\text{Diag}\left(\mathbf{h}_{n}^{\text{bs}}\right)\mathbf{h}_{n,k}^{p}$. Moreover, we define $\mathbf{U}_n^{p}=\mathbf{u}_n^{p}\left(\mathbf{u}_n^{p}\right)^H$, $\forall {p}\in \{{\text{r}},{\text{t}}\}$, which satisfy $\mathbf{U}_n^{p}\succeq 0$, ${\text{Rank}}(\mathbf{U}_n^{p})=1$, and ${\text{Diag}}\left(\mathbf{U}_n^{p}\right)=\boldsymbol{\beta}_n^{p}$, where $\boldsymbol{\beta}_n^{p}\triangleq\left[\beta_{n,1}^{p},\cdots,\beta_{n,m}^{p},\cdots,\beta_{n,M}^{p}\right]$, $\forall {p}\in \{\text{r},\text{t}\}$, $\beta_{n,m}^{\text{r}}=\alpha_{n,m}\varsigma_{n,m}^2$ and $\beta_{n,m}^{\text{t}}=\alpha_{n,m}\left(1-\varsigma_{n,m}^2\right)$. We denote $\mathbf{U}^{\text{r}}=\{\mathbf{U}_n^{\text{r}}, \forall n\}$ and $\mathbf{U}^{\text{t}}=\{\mathbf{U}_n^{\text{t}}, \forall n\}$ as the beamforming matrices. Subsequently, given the fixed UAV trajectory and the power allocation coefficients, the optimization problem \eqref{eq: optimization_problem} can be reformulated as

\begin{subequations}
\label{eq: optimization_problem of Beamforming A}
\begin{equation}
\label{eq: MSE_objective_1}
    \max_{\left\{\mathbf{U}^{\text{r}}, \mathbf{U}^{\text{t}} \right\}} \sum_{n=1}^{N} \sum_{k\in\mathcal{K}} R_{n,k}^{p}
\end{equation}
\begin{equation}
    \label{eq: STAR-RIS coefficients}
    \begin{split}
   {\rm{s.t.}} \   &\beta_{n,m}^{\text{r}}\geq 0, \beta_{n,m}^{\text{t}}\geq 0,\\
    \beta_{n,m}^{\text{r}}&+\beta_{n,m}^{\text{t}}=\alpha_{n,m}\geq1, \forall{m}, n,
    \end{split}
\end{equation}
\begin{equation}
\label{eq: A_element-wise power constraint of subproblem}
  P_n^\text{max}\alpha_{n,m}|h_{n,m}^{\text{bs}}|^{2}+\alpha_{n,m}\sigma_{\text{v}}^{2}\leq p_{n,m}^{\text{max}}, \forall{m}, n,
\end{equation}
\begin{equation}
\label{eq: B_total power constraint of subproblem}
  P_n^\text{max}||\mathbf{A}_n\mathbf{h}_{n}^{\text{bs}}||_{2}^{2}+\sigma_{\text{v}}^{2} ||\mathbf{A}_n||_{\text{F}}^{2}\leq P_{n}^{\text{max}},\forall n,
\end{equation}
\begin{equation}
\label{eq: rank one constraint}
    \text{Rank}\left(\mathbf{U}_n^{p}\right)=1, \forall {p}\in\{\text{r},\text{t}\}, n,
\end{equation}
\begin{equation}
\label{eq: diagonal constraint}
    \text{Diag}\left(\mathbf{U}_n^{p}\right)=\boldsymbol{\beta}_n^{p}, \forall{p}\in\{\text{r},\text{t}\}, n,
\end{equation}
\begin{equation}
\label{eq: semidefinite constraint}
    \mathbf{U}_n^{p}\succeq0, \forall {p}\in\{\text{r},\text{t}\}, n,
\end{equation}
\begin{equation}
\label{eq: decoding order constraint of channel}
    \begin{split}
        \text{Tr}&\left(\mathbf{U}_n^{p}\mathbf{H}_{n,k}^{\text{bg}-p}\right)\geq\text{Tr}\left(\mathbf{U}_n^{p}\mathbf{H}_{n,l}^{\text{bg}-{p}}\right),\\
        &\text{if}\  \mu_n(k) > \mu_n(l), \forall {p}\in\{\text{r},\text{t}\},
    \end{split}
\end{equation}
\begin{equation}
\label{eq: 19 constraints of former}
    \eqref{eq: reflected & transmitted coefficient constraint},  
\end{equation}
\end{subequations}
where $\mathbf{A}_n=\mathbf{U}_n^{\text{r}}+\mathbf{U}_n^{\text{t}}$, $\mathbf{H}_{n,k}^{\text{bg}-{p}}=\mathbf{h}_{n,k}^{\text{bg}-{p}}\left(\mathbf{h}_{n,k}^{\text{bg}-{p}}\right)^H$ and $\mathbf{H}_{n,k}^{p}=\mathbf{h}_{n,k}^{p}\left(\mathbf{h}_{n,k}^{p}\right)^H$. 

To address the non-convex problem \eqref{eq: optimization_problem of Beamforming A}, we first introduce the slack vectors $\mathbf{c}_n=\left[c_{n,1},\cdots,c_{n,K}\right]^T$, $\mathbf{d}_n=\left[d_{n,1},\cdots,d_{n,K}\right]^T$, where $c_{n,k}$ and $d_{n,k}$ are defined as 
\begin{equation}
\label{eq: receiver signal power}
   \frac{1}{c_{n,k}}=\text{Tr}\left(\mathbf{U}_n^{p}\mathbf{H}_{n,k}^{\text{bg}-{p}}\right)p_{n,k}, \forall {p}\in\{\text{r}, \text{t}\},
\end{equation}
\begin{equation}
   \begin{split} 
\label{eq: interference signal power}
   d_{n,k}= & \text{Tr}\left(\mathbf{U}_n^{p}\mathbf{H}_{n,k}^{\text{bg}-{p}}\right)
   \sum\limits_{\mu(i)>\mu(k)} p_{n,i} \noindent \\ 
   &+\text{Tr}\left(\mathbf{U}_n^{p}\mathbf{H}_{n,k}^{p}\right)\sigma_{\text{v}}^2+\sigma_{n,k}^2, \forall {p}\in\{\text{r}, \text{t}\},
    \end{split}
\end{equation}
respectively. A such, the achievable rate at IoT device $k$ can be represented as 
\begin{equation}
\label{eq:achievable rate of variable replacement}
   R_{n,k}^{p}=\log_2\left(1+\frac{1}{c_{n,k}d_{n,k}}\right), \forall {p}\in \{\text{r},\text{t}\}.
\end{equation}
We can utilize the first-order taylor expansion to derive the locally lower bound of~\eqref{eq:achievable rate of variable replacement}, as shown in \eqref{eq: first-order Taylor of Beamforming objective function} at the top of this page, where $c_{n,k}^{\tau_1}$ and $d_{n,k}^{\tau_1}$ are the given local points of $c_{n,k}$ and $d_{n,k}$, respectively, in the $\tau_1$-th iteration of the SCA. Moreover, the slack variables $\mathbf{C}\triangleq\{\mathbf{c}_n, \forall n\}$ and $\mathbf{D}\triangleq\{\mathbf{d}_n, \forall n\}$ are introduced. Then, by using the convex lower bound \eqref{eq: first-order Taylor of Beamforming objective function} to replace \eqref{eq: MSE_objective_1}, the optimization problem \eqref{eq: optimization_problem of Beamforming A} can be reformulated as 

\setcounter{equation}{18}
\begin{subequations}
\label{eq: optimization_problem of Beamforming B}
\begin{equation}
\label{eq: MSE_objective}
    \max_{\left\{\mathbf{U}^{\text{r}}, \mathbf{U}^{\text{t}}, \mathbf{C}, \mathbf{D} \right\}} \sum_{n=1}^{N} \sum_{k\in\mathcal{K}} \tilde{R}_{n,k}^{p} 
\end{equation}
\begin{equation}
    \label{eq: receiver signal power constraint}
    {\rm{s.t.}} \ \frac{1}{c_{n,k}}\leq\text{Tr}\left(\mathbf{U}_n^{p} \mathbf{H}_{n,k}^{\text{bg}-{p} }\right)p_{n,k}, \forall {p} \in\{\text{r},\text{t}\}, \forall n, k,
\end{equation}
\begin{equation}
\label{eq: interference signal power constraint}
    \begin{split}
   d_{n,k}\geq &\text{Tr}\left(\mathbf{U}_n^{p} \mathbf{H}_{n,k}^{\text{bg}-{p} }\right)
   \sum\limits_{\mu(i)>\mu(k)} p_{n,i} \\
   &+\text{Tr}\left(\mathbf{U}_n^{p} \mathbf{H}_{n,k}^{p} \right)\sigma_{\text{v}}^2+\sigma_{n,k}^2,
   \forall {p} \in\{\text{r},\text{t}\}, n, k,
    \end{split}
\end{equation}
\begin{equation}
\label{eq: 24 constraints of former}
      \eqref{eq: reflected & transmitted coefficient constraint}, \eqref{eq: STAR-RIS coefficients}\sim \eqref{eq: decoding order constraint of channel}.
\end{equation}
\end{subequations}

Now, the only non-convexity left in problem~\eqref{eq: optimization_problem of Beamforming B} is the rank-one constraint \eqref{eq: rank one constraint}. To deal with this issue, we first transform \eqref{eq: rank one constraint}  into the equality constraint as below:
\begin{equation}
\label{eq: rank-one constraint equation 1}
   \zeta_n^{p} \triangleq\text{Tr}\left(\mathbf{U}_n^{p} \right)-||\mathbf{U}_n^{p} ||_2=0, \forall {p} \in \{\text{r},\text{t}\},
\end{equation}
where $||\mathbf{U}_n^{p} ||_2=\sigma_1\left(\mathbf{U}_n^{p} \right)$ denotes the spectral norm, with $\sigma_1(\mathbf{U}_n^{p} )$ representing the largest singular value of the matrix $\mathbf{U}_n^{p} $. Note that for any $\mathbf{U}_n^{p} $, the inequality $\text{Tr}(\mathbf{U}_n^{p} )-||\mathbf{U}_n^{p} ||_2\geq0$ always holds, with equality occurring if and only if $\mathbf{U}_n^{p} $ is a rank-one matrix. Hence, the equality constraint \eqref{eq: rank-one constraint equation 1} is satisfied exclusively when ${\mathbf{U}_n^{p} }$ is rank-one. 

Next, we apply a penalty-based method to tackle problem \eqref{eq: optimization_problem of Beamforming B}. By adding the quality constraint \eqref{eq: rank-one constraint equation 1} into the objective function as a penalty term, problem \eqref{eq: optimization_problem of Beamforming B} can be expressed in detail as
\begin{subequations}
\label{eq: optimization_problem of Beamforming C}
\begin{equation}
\label{eq: MSE_objective}
    \max_{\left\{\mathbf{U}^{\text{r}}, \mathbf{U}^{\text{t}},\mathbf{C}, \mathbf{D} \right\}} \sum_{n=1}^{N}\left[\sum\limits_{k\in\mathcal{K}}\tilde{R}_{n,k}^{p} -\eta \sum\limits_{p\in\{\text{r},\text{t}\}} \zeta_n^p \right]
\end{equation}
\begin{equation}
\label{eq: 26 constraints of former}
    {\rm{s.t.}} \  \eqref{eq: reflected & transmitted coefficient constraint}, \eqref{eq: STAR-RIS coefficients}\sim \eqref{eq: decoding order constraint of channel},  \eqref{eq: receiver signal power constraint}\sim \eqref{eq: interference signal power constraint},
\end{equation}
\end{subequations}
where equality constraint \eqref{eq: rank-one constraint equation 1} is relaxed and added to the objective function as a penalty term.Here, $\eta>0$ denotes the penalty factor that adds a penalty to the objective function whenever $\{\mathbf{U}_n^{p} \}$ is not a rank-one matrix. It can be demonstrated that as $\eta \rightarrow \infty$ (or becomes sufficiently large), the solution $\{\mathbf{U}_n^{p} \}$ of problem \eqref{eq: optimization_problem of Beamforming C} will comply with the equality constraint \eqref{eq: rank-one constraint equation 1}. However, if $\eta$ is initially set too large, the penalty term will dominate the objective function \eqref{eq: optimization_problem of Beamforming C}, diminishing the influence of the desired maximum sum rate of IoT devices on the optimization result. To prevent this, we begin by setting $\eta$ to a relatively small value to determine an appropriate initial point. Then $\eta$ is gradually increased until it reaches a suitable value to ensure acceptable rank-one matrices.

The problem \eqref{eq: optimization_problem of Beamforming C} is still a non-convex optimization problem due to the penalty term. To address this, we replace $||\mathbf{U}_n^{p} ||_2$ with its linear lower bound utilizing the first-order Taylor expansion. Specifically, in the $\tau_1$-th iteration of the SCA, with the given point $\mathbf{U}_n^{{p} (\tau_1)}$, the convex upper bound of the penalty term can be derived as follows:
\begin{equation}
\label{eq: rank-one constraint equation 2}
   \text{Tr}(\mathbf{U}_n^{p} )-||\mathbf{U}_n^{p} ||_2\leq \text{Tr}(\mathbf{U})_n^{p} -\bar{\mathbf{U}}_n^{{p} (\tau_1)}\triangleq\zeta_n^{{p} (\tau_1)}, 
\end{equation}
where $\bar{\mathbf{U}}_n^{{p} (\tau1)}\triangleq||\mathbf{U}_n^{{p} (\tau_1)}||_2+\text{Tr}\left[\bar{\mathbf{x}}(\mathbf{U}_n^{{p} (\tau_1)})(\bar{\mathbf{x}}(\mathbf{U}_n^{{p} (\tau_1)}))^H(\mathbf{U}_n^{p} -\mathbf{U}_n^{{p} (\tau_1)})\right]$ and $\bar{\mathbf{x}}(\mathbf{U}_n^{{p} (\tau_1)})$ denotes the eigenvector associated with the largest eigenvalue of the matrix $\mathbf{U}_n^{{p} (\tau_1)}$. 

Subsequently, problem \eqref{eq: optimization_problem of Beamforming C} can be equivalently addressed by solving the following problem
\begin{subequations}
\label{eq: optimization_problem of Beamforming D}
\begin{equation}
\label{eq: MSE_objective}
    \max_{\left\{\mathbf{U}^{\text{r}}, \mathbf{U}^{\text{t}},\mathbf{C}, \mathbf{D} \right\}} \sum_{n=1}^{N}\left[\sum\limits_{k\in\mathcal{K}}\tilde{R}_{n,k}^{p}-\eta \sum\limits_{p\in\{\text{r},\text{t}\}} \zeta_n^{p(\tau1)} \right]
\end{equation}
\begin{equation}
\label{eq: 28 former constraint}
    {\rm{s.t.}} \  \eqref{eq: reflected & transmitted coefficient constraint}, \eqref{eq: STAR-RIS coefficients}\sim \eqref{eq: B_total power constraint of subproblem}, \eqref{eq: diagonal constraint}\sim \eqref{eq: decoding order constraint of channel}, \eqref{eq: receiver signal power constraint}\sim \eqref{eq: interference signal power constraint}.
\end{equation}
\end{subequations}
The relaxed problem \eqref{eq: optimization_problem of Beamforming D}, structured as a standard convex semi-definite program (SDP) can be effectively addressed by using optimization tools like CVX~\cite{boyd2004convex}. 

Note that the proposed algorithm for solving the original problem~\eqref{eq: optimization_problem of Beamforming A} consists of two loops. Specifically, in the outer loop, the penalty factor is progressively increased with each iteration, following the update rule: $\eta=\omega\eta$, where $\omega >1$. The termination of the algorithm occurs when the penalty term meets the following criterion:
\begin{equation}
\label{eq: rank-one constraint equation 3}
\text{max}\left\{\text{Tr}(\mathbf{U}_n^{p} )-||\mathbf{U}_n^{p} ||_2, \forall {p}  \in \{\text{r}, \text{t}\}\right\}\leq \varepsilon, 
\end{equation}
where $\varepsilon>0$ represents a predefined threshold for the highest allowable deviation from the equality constraint \eqref{eq: rank-one constraint equation 1}. Consequently, as $\eta$ increases, constraint \eqref{eq: rank-one constraint equation 1} will be met with a error of $\varepsilon$. With the specified penalty factor, in the inner loop, the variable $\{\mathbf{U}^{\text{r}}, \mathbf{U}^{\text{t}}\}$ are updated by solving the relaxed form of problem \eqref{eq: optimization_problem of Beamforming D}. The relaxed problem \eqref{eq: optimization_problem of Beamforming D} exhibits a monotonically decreasing objective function value with each inner loop iteration, while remaining bounded below. Consequently, the penalty-based iterative method ensures convergence to a feasible point of the original problem \eqref{eq: optimization_problem of Beamforming A} as $\eta$ grows infinitely large. The specific steps of the proposed algorithm are outlined in \textbf{Algorithm 1}.

\begin{algorithm}
	\renewcommand{\algorithmicrequire}{\textbf{Input:}}
	\renewcommand{\algorithmicensure}{\textbf{Output:}}
	\caption{Proposed penalty-based SCA algorithm for solving problem \eqref{eq: optimization_problem}}
	\label{alg1: Active STAR-RIS's Beamforming Vectors}
	\begin{algorithmic}[1]
		\State Initialize the feasible points $\left\{\mathbf{U}^{\text{r}(0)},\mathbf{U}^{\text{t}(0)}, \mathbf{C}^{(0)}, \mathbf{D}^{(0)} \right\}$ and the penalty factor $\eta$;
		\Repeat
            \State Set $\tau_1=0$;
                \Repeat
		      \State With the given points $\left\{\mathbf{U}^{\text{r}(\tau_1)}, \mathbf{U}^{\text{t}(\tau_1)}, \mathbf{C}^{(\tau_1)}, \mathbf{D}^{(\tau_1)}\right\}$, solve the relaxed problem (28);
		      \State Update $\left\{\mathbf{U}^{\text{r}(\tau_1)}, \mathbf{U}^{\text{t}(\tau_1)}, \mathbf{C}^{(\tau_1)}, \mathbf{D}^{(\tau_1)}\right\}$;
            \State $\tau_1=\tau_1+1$;
		\Until the penalty term value is below a predefined threshold $\varepsilon>0$ \textbf{or} $\tau_1$ reaches the inner maximum number $\tau_1^{\text{max}}$;
		\State Update $\left\{\mathbf{U}^{\text{r}(0)},\mathbf{U}^{\text{t}(0)}, \mathbf{C}^{(0)}, \mathbf{D}^{(0)} \right\}$ with the inner loop solutions $\left\{\mathbf{U}^{\text{r}(\tau_1)}, \mathbf{U}^{\text{t}(\tau_1)}, \mathbf{C}^{(\tau_1)}, \mathbf{D}^{(\tau_1)} \right\}$;
            \State Update $\eta=\omega\eta$;
        \Until the penalty term value is below a predefined threshold $\varepsilon$.
	\end{algorithmic}  
\end{algorithm}

\setcounter{TempEqCnt}{\value{equation}} 
\setcounter{equation}{26} 
\begin{figure*}[ht]
    \begin{equation}
    \label{eq:k-th transmitted or reflection use}
           R_{n,k}^{p} =\text{log}_2\left(1+\frac{\text{Tr}\left(\mathbf{U}_n^{p} \bar{\mathbf{H}}_{n,k}^{\text{bg}-{p} }\right)p_{n,k}}{\text{Tr}\left(\mathbf{U}_n^{p} \bar{\mathbf{H}}_{n,k}^{\text{bg}-{p} }\right) \sum\limits_{\mu(i)>\mu(k)} p_{n,i} +l_n^{\tau_0}\text{Tr}\left(\mathbf{U}_n^{p} \bar{\mathbf{H}}_{n,k}^{p} \right)\sigma_{\text{v}}^2+l_n^{\tau_0}j_{n,k}^{\tau_0}\sigma_{n,k}^2}\right),
    \end{equation}
    \hrulefill
\end{figure*}

\subsection{UAV's Trajectory Design}
\label{sec:coordinate-transformation}
With the active STAR-RIS amplification gain and reflection/transmission matrices $\mathbf{U}^{\text{r}}$ and $\mathbf{U}^{\text{t}}$, and the power allocation matrix $\mathbf{P}$ fixed, problem \eqref{eq: optimization_problem} can be reformulated as

\setcounter{equation}{24}
\begin{subequations}
\label{eq:optimization_problem of UAV Trajectory A}
\begin{equation}
\label{eq:MSE_objective}
    \max_{\left\{\mathbf{Q} \right\}} \sum_{n=1}^{N} \sum_{k\in\mathcal{K}} R_{n,k}^{p}  
\end{equation}
\begin{equation}
\label{eq: 30 former constraint}
    {\rm{s.t.}} \ \ \eqref{eq: A_trajectory} \sim \eqref{eq: C_trajectory}, \eqref{eq: A_element-wise power constraint}, \eqref{eq: B_total power constraint}, \eqref{eq: decoding strategy of channel gain}.
\end{equation}
\end{subequations}
However, the problem \eqref{eq:optimization_problem of UAV Trajectory A} is a non-convex optimization problem with respect to the UAV trajectory matrix $\mathbf{Q}$. For solving this problem, we start by expanding the expression of $R_{n,k}^{p} $ to illustrate its dependency on $\mathbf{Q}$. Specifically, the term $|g_{n,k}|^2$ can be rewritten as
\begin{equation}
\label{eq: |g_{n,k}|^2 another mode}
   |g_{n,k}|^2=\frac{\text{Tr}\left(\mathbf{U}_n^{p}\bar{\mathbf{H}}_{n,k}^{\text{bg}-{p}}\right)}{d_{\text{bs}}(n)^{\tau_0}d_{\text{s}k}(n)^{\tau_0}}, \forall {p}\in\{{\text{r}},{\text{t}}\},
\end{equation}
where $\bar{\mathbf{H}}_{n,k}^{\text{bg}-{p}}=\bar{\mathbf{h}}_{k}^{\text{bg}-{p}}\left(\bar{\mathbf{h}}_{n,k}^{\text{bg}-p}\right)^H$, $\bar{\mathbf{h}}_{n,k}^{\text{bg}-{p}}=\text{Diag}\left(\bar{\mathbf{h}}_{n}^{\text{bs}}\right)\bar{\mathbf{h}}_{n,k}^{p}$ with $\bar{\mathbf{h}}_{n,k}^{p}=\mathbf{h}_{n,k}^{p}\sqrt{d_{\text{s}k}(n)^{\tau_0}}$ and $\bar{\mathbf{h}}_{n}^{\text{bs}}=\mathbf{h}_{n}^{\text{bs}}\sqrt{d_{\text{bs}}(n)^{\tau_0}}$, are constants that are irrelevant to the variable $\mathbf{q}_u(n)$.

Then, the slack variables $\mathbf{L}\triangleq\left\{l_n, \forall n\right\}$, $l_n\geq d_{\text{bs}}(n)=||\mathbf{q}_{\text{B}}-\mathbf{q}_{\text{u}}(n)||$ and $\mathbf{J}\triangleq\left\{\mathbf{j}_{n}, \forall n\right\}$, $\mathbf{j_n}\triangleq\left\{j_{n,k}, \forall k\right\}$, $j_{n,k}\geq d_{\text{s}k}(n)=||\mathbf{q}_{\text{u}}(n)-\mathbf{q}_k||$ are introduced, and the data rate of the $k$-th IoT device in the $n$-th time slot can be reformulated as \eqref{eq:k-th transmitted or reflection use}, which is displayed at the top of the next page. To this end, the optimization problem \eqref{eq:optimization_problem of UAV Trajectory A} can be reformulated as

\setcounter{equation}{27}
\begin{subequations}
\label{eq:optimization_problem of UAV Trajectory B}
\begin{equation}
\label{eq:MSE_objective}
    \max_{\left\{\mathbf{Q}, \mathbf{L}, \mathbf{J} \right\}} \sum_{n=1}^{N} \sum_{k\in\mathcal{K}} R_{n,k}^{p}
\end{equation}
\begin{equation}
\label{eq: BS to UAV distance}
    {\rm{s.t.}} \ \ l_n\geq ||\mathbf{q}_{\text{B}}-\mathbf{q}_{\text{u}}(n)||, \forall n,
\end{equation}
\begin{equation}
\label{eq: UAV to IoT device distance}
    j_{n,k}\geq \||\mathbf{q}_{\text{u}}(n)-\mathbf{q}_k||, \forall n, k,
\end{equation}
\begin{equation}
\label{eq: 33 former constraint}
    \eqref{eq: A_trajectory} \sim \eqref{eq: C_trajectory}, \eqref{eq: A_element-wise power constraint}, \eqref{eq: B_total power constraint}, \eqref{eq: decoding strategy of channel gain}.
\end{equation}
\end{subequations}
The problem \eqref{eq:optimization_problem of UAV Trajectory A} and \eqref{eq:optimization_problem of UAV Trajectory B} are equivalent if \eqref{eq: BS to UAV distance} and \eqref{eq: UAV to IoT device distance} hold equality. Nevertheless, problem \eqref{eq:optimization_problem of UAV Trajectory B} is still a non-convex problem. According to the Taylor's theorem, we obtain the first-order taylor approximation of $R_{n,k}^{p}$, $|g_{n,k}|^2$, $l_n^2$, $j_{kn}^2$, in the $\tau_2$-th iteration of the SCA method as follows:
\begin{equation}
\label{eq: first-order Taylor expansion of data rate of reflected or transmitted IoT device k}
    \begin{split}
    R_{n,k}^{p} \geq&  \check{R}_{n,k(\mathbf{L},\mathbf{J})}^{p}\\
    =&\text{log}_2\left(A_{n,k}^{(\tau_2)}\right)-B_{n,k}^{(\tau_2)}C_{n,k}^{(\tau_2)}\left(l_n-l_n^{(\tau_2)}\right)\\
    &-B_{n,k}^{(\tau_2)}D_{n,k}^{(\tau_2)}\left(j_{n,k}-j_{n,k}^{(\tau_2)}\right), \forall p\in\{\text{r},\text{t}\},
   \end{split}
\end{equation}

\begin{equation}
\label{eq: |g_{n,k}|^2 first-order Taylor}
    \begin{split}
    |g_{n,k}|^2\geq& \frac{\text{Tr}\left(\mathbf{U}_n^{p}\bar{\mathbf{H}}_{n,k}^{\text{bg}-{p}}\right)}{l(n)^{\tau_0(\tau_2)}j_{n,k}(n)^{\tau_0(\tau_2)}}\\
    & -\tau_0\frac{\text{Tr}\left(\mathbf{U}_n^p\bar{\mathbf{H}}_{n,k}^{\text{bg}-{p}}\right)}{l(n)^{\left(\tau_0+1\right)(\tau_2)}j_{n,k}(n)^{\tau_0(\tau_2)}}\left(l(n)-l(n)^{(\tau_2)}\right)\\
    & -\tau_0\frac{\text{Tr}\left(\mathbf{U}_n^p\bar{\mathbf{H}}_{n,k}^{\text{bg}-{p}}\right)}{l(n)^{(\tau_0)(\tau_2)}j_{n,k}(n)^{\left(\tau_0+1\right)(\tau_2)}}\left(j_{n,k}(n)-j_{n,k}(n)^{(\tau_2)}\right)\\
    =&O_{n,k}, \forall p\in\{\text{r},\text{t}\},
    \end{split}
\end{equation}

\begin{equation}
\label{eq: first-order Taylor expansion of l_n}
    l_n^2\geq 2l_n^{(\tau_2)}l_n-\left(l_n^{(\tau_2)}\right)^2,
\end{equation}

\begin{equation}
\label{eq: first-order Taylor expansion of j_kn}
    j_{n,k}^2\geq 2j_{n,k}^{(\tau_2)}j_{n,k}-\left(j_{n,k}^{(\tau_2)}\right)^2,
\end{equation}
where $A_{n,k}^{(\tau_2)}$, $B_{n,k}^{(\tau_2)}$, $C_{n,k}^{(\tau_2)}$, $D_{n,k}^{(\tau_2)}$ and $E_{n,k}^{(\tau_2)}$ can be expressed as follows

\begin{equation}
\label{eq: A after iteration}
   A_{n,k}^{(\tau_2)}=1+\frac{p_{n,k}}{L_{n,k}},
\end{equation}

\begin{equation}
\label{eq: B after iteration}
    B_{n,k}^{(\tau_2)}=\frac{p_{n,k}\text{log}_2 e}{(p_{n,k} +L_{n,k})(L_{n,k})},
\end{equation}

\begin{equation}
\label{eq: C after iteration}
    C_{n,k}^{(\tau_2)}=\tau_0l_n^{(\tau_0-1)(\tau_2)}\left( \sigma_{\text{v}}^2N_{n,k}^{p}+j_{n,k}^{\tau_0(\tau_2)}\sigma_{n,k}^2M_{n,k}^{p}\right),
\end{equation}

\begin{equation}
\label{eq: D after iteration}
    D_{n,k}^{(\tau_2)}=\tau_0l_n^{\tau_0(\tau_2)}j_{n,k}^{(\tau_0-1)(\tau_2)}\sigma_{n,k}^2M_{n,k}^{p},
\end{equation}  
where $p=\{\text{r},\text{t}\}$, $L_{n,k}=\sum_{\mu(i)>\mu(k)} p_{n,i} +l_n^{\tau_0(\tau_2)}\sigma_{\text{v}}^2N_{n,k}^{p}+l_n^{\tau_0(\tau_2)}j_{n,k}^{\tau_0(\tau_2)}\sigma_{n,k}^2M_{n,k}^{p}$, $N_{n,k}^{p}=\frac{\text{Tr}\left(\mathbf{U}_n^{p}\bar{\mathbf{H}}_{n,k}^{p}\right)}{\text{Tr}\left(\mathbf{U}_n^{p}\bar{\mathbf{H}}_{n,k}^{\text{bg}-{p}}\right)}$, and $M_{n,k}^{p}=\frac{1}{\text{Tr}\left(\mathbf{U}_n^{p}\bar{\mathbf{H}}_{n,k}^{\text{bg}-{p}}\right)}$. Replacing $R_{n,k}^{p}$ by $\check{R}_{n,k}^{p}$, the optimization problem for the UAV trajectory design can be reformulated as
\begin{subequations}
\label{eq:optimization_problem of UAV Trajectory C}
\begin{equation}
\label{eq:MSE_objective}
    \max_{\left\{\mathbf{Q}, \mathbf{L}, \mathbf{J} \right\}} \sum_{n=1}^{N} \sum_{k\in\mathcal{K}} \check{R}_{n,k(\mathbf{L},\mathbf{J})}^{p} 
\end{equation}
\begin{equation}
\label{eq:the quality of channel}
     {\rm{s.t.}} \ O_{n,k}\geq O_{n,j}, \text{if} \ \mu_n(k) > \mu_n(j), \forall n,
\end{equation}
\begin{equation}
\label{eq:BS to UAV distance constraint}
    ||\mathbf{q}_{\text{B}}-\mathbf{q}_{\text{u}}(n)||^2+\left(l_n^{(\tau_2)}\right)^2-2l_n^{(\tau_2)}l_n\leq 0, \forall n,
\end{equation}
\begin{equation}
\label{eq:UAV to IoT device distance constraint}
    ||\mathbf{q}_{\text{u}}(n)-\mathbf{q}_k||^2+\left(j_{n,k}^{(\tau_2)}\right)^2-2j_{n,k}^{(\tau_2)}j_{n,k}\leq 0, \forall n, k,
\end{equation}
\begin{equation}
\label{eq: 42 former constraint}
   \eqref{eq: A_trajectory} \sim \eqref{eq: C_trajectory}, \eqref{eq: A_element-wise power constraint}, \eqref{eq: B_total power constraint}.
\end{equation}
\end{subequations}
Problem (42) can be addressed efficiently using convex optimization tools like CVX. For clarity, the algorithm developed for solving the UAV trajectory optimization problem is outlined in \textbf{Algorithm 2}.

\begin{algorithm}
	\renewcommand{\algorithmicrequire}{\textbf{Input:}}
	\renewcommand{\algorithmicensure}{\textbf{Output:}}
	\caption{Proposed SCA-based algorithm for solving \eqref{eq:optimization_problem of UAV Trajectory C}}
	\label{alg1: Power Allocation}
	\begin{algorithmic}[1]
		\State Initialize feasible point $\mathbf{Q}^{(0)}$ and threshold value $\delta$; \\ Set $\tau_2=0$;
		\Repeat
		\State Optimize $\mathbf{Q}$ by solving the problem (42);
		\State Update $\left\{\mathbf{Q}^{(\tau_2+1)}\right\}$ with current solution $\{\mathbf{Q}\}$;
            \State  $\tau_2=\tau_2+1$;
		\Until the fractional increase of the objective value is below $\delta$;
		\State Return the optimal solution $\mathbf{Q}^*$.
	\end{algorithmic}  
\end{algorithm}

\vspace{-0.2cm}
\subsection{Power Allocation Optimization}
\label{sec:coordinate-transformation}
When the active STAR-RIS beamforming matrices $\mathbf{U}^{\text{r}}$ and $\mathbf{U}^{\text{t}}$, and the trajectory matrix $\mathbf{Q}$ of the UAV are fixed, problem \eqref{eq: optimization_problem} can be rewritten as
\begin{subequations}
\label{eq:optimization_problem of power allocation A}
\begin{equation}
\label{eq:MSE_objective}
    \max_{\left\{\mathbf{P} \right\}} \sum_{n=1}^{N} \sum_{k\in\mathcal{K}} R_{n,k}^{p} 
\end{equation}
\begin{equation}
\label{eq: 43 former constraint}
    {\rm{s.t.}} \ \ \eqref{eq: power allocation by decoding order}, \eqref{eq: total power allocation}.
\end{equation}
\end{subequations}
From the previous description, it is known that $\mu(k)=k$ represents the decoding order. We define $\varrho_{n,k}=
   \sum_{i=k}^K p_{n,i} $, $\forall k\in\mathcal{K}$. For IoT device $k$, the achievable data rate given in \eqref{eq: data rate of reflected IoT device i} can be reformulated as

\begin{equation}
\label{eq: achievable rate at IoT device k of power allocation}
   \begin{split}
    R_{n,k}^{p}=&\text{log}_2\left(|g_{n,k}|^{2}
   \sum_{i=k}^K p_{n,i} +\sigma_{\text{v}}^2||\mathbf{A}_n\mathbf{E}_{n}^{p}\mathbf{h}_{n,k}^{p}||_2^2+\sigma_{n,k}^2\right)\\
    &-\text{log}_2\left(|g_{n,k}|^{2}\sum_{i>k}^K p_{n,i}+\sigma_{\text{v}}^2||\mathbf{A}_n\mathbf{E}_{n}^{p}\mathbf{h}_{n,k}^{p}||_2^2+\sigma_{n,k}^2\right)\\
    =&\text{log}_2\left(|g_{n,k}|^{2}\varrho_{n,k}+\sigma_{\text{v}}^2||\mathbf{A}_n\mathbf{E}_{n}^{p}\mathbf{h}_{n,k}^{p}||_2^2+\sigma_{n,k}^2\right)\\
    &-\text{log}_2\left(|g_{n,k}|^{2}\varrho_{n,k+1}+\sigma_{\text{v}}^2||\mathbf{A}_n\mathbf{E}_{n}^{p}\mathbf{h}_{n,k}^{p}||_2^2+\sigma_{n,k}^2\right),
   \end{split}
\end{equation}
where $p\in\{\text{r},\text{t}\}$, and $\varrho_{n,K+1}\triangleq 0$. Thus, the optimization problem \eqref{eq:optimization_problem of power allocation A} can be expressed as follows:
\begin{subequations}
\label{eq:optimization_problem power allocation B}
\begin{equation}
\label{eq:MSE_objective}
    \max_{\left\{\varrho_{n,k} \right\}} \sum_{n=1}^{N} \sum_{k\in\mathcal{K}} R_{n,k}^{p}
\end{equation}
\begin{equation}
\label{eq:power allocation each IoT device}
     {\rm{s.t.}} \ \ \varrho_{n,1}-\varrho_{n,2}\geq \varrho_{n,2}-\varrho_{n,3}\geq ...\geq \varrho_{n,K}\geq0, \forall n,
\end{equation}
\begin{equation}
\label{eq:power allocation of all IoT devices}
    \varrho_{n,1}\leq P_\text{B}^\text{max}, \forall n.
\end{equation}
\end{subequations}
However, due to the non-concave objective function, the problem \eqref{eq:optimization_problem power allocation B} is challenging to solve. As shown above, the objective function takes the form of the difference of two concave functions. With the fixed local point $\varrho_{n, k+1}^{(\tau_3)}$ during the $\tau_3$-th iteration of the SCA method, we utilize the first-order Taylor expansion to derive a concave lower bound, as shown below:
\begin{equation}
\label{eq: first-order Taylor expansion}
      \begin{split}
    R_{n,k}^{p} \geq &\bar{R}_{n,k(\varrho_{n, k})}^{p}\\ =&\text{log}_2\left(|g_{n,k}|^{2}\varrho_{n,k}+\sigma_{\text{v}}^2||\mathbf{A}_n\mathbf{E}_{n}^{p}\mathbf{h}_{n,k}^{p}||_2^2+\sigma_{n,k}^2\right)\\    
    &-\text{log}_2\left(|g_{n,k}|^{2}\varrho_{n,k+1}^{(\tau_3)}+\sigma_{\text{v}}^2||\mathbf{A}_n\mathbf{E}_{n}^{p}\mathbf{h}_{n,k}^{p}||_2^2+\sigma_{n,k}^2\right)\\
    &-\frac{|g_{n,k}|^{2}\left(\varrho_{n,k+1}-\varrho_{n,k+1}^{(\tau_3)}\right)\text{log}_2e}{|g_{n,k}|^{2}\varrho_{n,k+1}^{(\tau_3)}+\sigma_{\text{v}}^2||\mathbf{A}_n\mathbf{E}_{n}^{p}\mathbf{h}_{n,k}^{p}||_2^2+\sigma_{n,k}^2}.
   \end{split}
\end{equation}
We replace the original objective function in problem \eqref{eq:optimization_problem power allocation B} with the concave lower bound. As such, the optimization problem can be rewritten as
\begin{subequations}
\label{eq:optimization_problem power allocation C}
\begin{equation}
\label{eq:MSE_objective}
    \max_{\left\{\varrho_{n,k} \right\}} \sum_{n=1}^{N} \sum_{k\in\mathcal{K}} \bar{R}_{n,k(\varrho_{n,k})}^{p} 
\end{equation}
\begin{equation}
\label{eq:eq: 47 former constraints}
    {\rm{s.t.}} \ \  \eqref{eq:power allocation each IoT device}, \eqref{eq:power allocation of all IoT devices}.
\end{equation}
\end{subequations}
Now, as the problem \eqref{eq:optimization_problem power allocation C} is in the convex form, it can be efficiently addressed using optimization tools like CVX. Due to the substitution with the concave lower bound, the objective function values obtained for problem \eqref{eq:optimization_problem power allocation C} generally serve as lower bounds for problem \eqref{eq:optimization_problem power allocation B}. Upon solving problem \eqref{eq:optimization_problem power allocation C}, the power allocation can be determined as $p_{n,k}^*=\varrho_{n,k}^*-\varrho_{n,k+1}^*$, $\forall k\in \mathcal{K}$. The steps for solving problem \eqref{eq:optimization_problem power allocation C} are outlined in \textbf{Algorithm 3}.

\begin{algorithm}
	\renewcommand{\algorithmicrequire}{\textbf{Input:}}
	\renewcommand{\algorithmicensure}{\textbf{Output:}}
	\caption{Proposed SCA based algorithm for solving \eqref{eq:optimization_problem power allocation C}}
	\label{alg1: Power Allocation}
	\begin{algorithmic}[1]
		\State Initialize feasible point $\mathbf{P}^{(0)}$ and threshold value $\delta'$;\\ 
  Calculate $\left\{\varrho_{n, k}^{(0)}\right\}$; \\
  Set iterative index $\tau_3=0$;
		\Repeat
		\State Optimize $\{\varrho_{n, k}\}$ by solving the problem (47);
		\State Update $\left\{\varrho_{n, k}^{(\tau_3+1)}\right\}$ with current solution $\{\varrho_{n, k}\}$;
            \State $\tau_3=\tau_3+1$;
		\Until the fractional increase of the objective value is below $\delta'$;
		\State  Return the optimal solution $\{\varrho_{n, k}^*\}$; \\Calculate $\mathbf{P}^*$.
	\end{algorithmic}  
\end{algorithm}

\vspace{-0.2cm}
\subsection{Proposed IoT Devices Decoding Order Design Scheme}

Note that the subproblems mentioned above are solved with a given decoding order. It is obvious to expect that, the straightforward way for the optimal decoding order design is to exhaustively search through all the possible decoding sequences and select the one that provides the best solution. However, this strategy results in extremely high computational complexity, in the order of $\mathcal{O}(K!)$, making it impractical even for a moderate number of IoT devices. To overcome this challenge, we develop a low-complexity decoding order design scheme that leverages the distances between the active STAR-RIS and the IoT devices. Specifically, as all the IoT devices share the same BS-active STAR-RIS channel, the active STAR-RIS-device channel is the main part that differentiates the received power strength. As such, according to the initialized UAV trajectory, the IoT devices located closer to the UAV are assigned lower decoding priorities in each time slot. The rule is formalized as follows:
\begin{equation}
\label{eq: decoding order scheme}
    \mu_n(k) < \mu_n(l)\ \text{if} \ ||\mathbf{q}_u^{(0)}(n)-\mathbf{q}_l|| \leq ||\mathbf{q}_u^{(0)}(n)-\mathbf{q}_k||.
\end{equation}   
Although the decoding order design based on the initial UAV trajectory may narrow the optimization space of the beamforming, trajectory design, and power allocation into a constrained region, this approach significantly reduces the computational burden while still providing a reasonable approximation to the optimal decoding order.

\subsection{Convergence and Complexity Analysis of the Proposed AO Algorithm}
\label{sec:Proposed Algorithm, Convergence and Complexity}

Following the above discussions, the proposed AO-based iterative algorithm is utilized to address the original problem \eqref{eq: optimization_problem}. Specifically, The AO-based iterative algorithm starts by initializing a feasible solution, including the aforementioned three subproblems. In each iteration, the algorithm first calculates the achievable sum rate based on the current values of these variables. Then, it alternately optimizes the the active STAR-RIS amplification gain and reflection/transmission coefficient, UAV trajectory, and the power allocation coefficients by invoking \textbf{Algorithm 1}, \textbf{Algorithm 2}, and \textbf{Algorithm 3}, respectively. This iterative process ensures that each component is refined in turn, contributing to the overall improvement of the system performance. The iteration continues until the objective value converges, indicating that further improvements are minimal. Throughout this process, the solutions obtained from each iteration are recorded and used to recalculate the sum rate. The updated results are then compared with those from the previous iteration. If an improvement is observed, the new solutions are retained, otherwise, the previous solutions are preserved to maintain stability. The iteration process terminates when the relative change in the sum rate between fractional iterations increases below a predefined threshold $\delta"$ or when the maximum iteration number is reached. 

\begin{algorithm}
	\renewcommand{\algorithmicrequire}{\textbf{Input:}}
	\renewcommand{\algorithmicensure}{\textbf{Output:}}
	\caption{Proposed AO-based algorithm for solving problem \eqref{eq: optimization_problem}}
	\label{alg1: AO}
	\begin{algorithmic}[1]
		\State Initialize feasible points $\left\{\mathbf{U}^{\text{r}(0)},\mathbf{U}^{\text{t}(0)}, \mathbf{C}^{(0)}, \mathbf{D}^{(0)} \right\}$, $\mathbf{Q}^{(0)}$, $\mathbf{P}^{(0)}$. \\ Initialize threshold value $\delta''$. \\ Set iterative index $\tau=0$.
		\Repeat
		\State Calculate $R_{\text{sum}}^{(\tau)}=R_{\text{sum}}\left(\mathbf{U}^{\text{r}(\tau)},\mathbf{U}^{\text{t}(\tau)}, \mathbf{C}^{(\tau)}, \mathbf{D}^{(\tau)}, \mathbf{Q}^{(\tau)}, \mathbf{P}^{(\tau)}\right)$;
		\Repeat
            \State Update active STAR-RIS beamforming matrices $\left\{\mathbf{U}^{\text{r}}, \mathbf{U}^{\text{t}}, \mathbf{C}, \mathbf{D}\right\}$ via \textbf{Algorithm 1};
            \State Update UAV trajectory matrix $\mathbf{Q}$ via \textbf{Algorithm 2};
         \State Update the power allocation matrix $\mathbf{P}$ via \textbf{Algorithm 3}.
		\Until The objective value of problem \eqref{eq: optimization_problem} converges.
            \State Record the obtained solution $\left\{\mathbf{U}^{\text{r}},\mathbf{U}^{\text{t}}, \mathbf{C}, \mathbf{D}\right\}, \mathbf{Q}, \mathbf{P}$;
		\State Calculate $R_{\text{sum}}=R_{\text{sum}}\left(\mathbf{U}^{\text{r}},\mathbf{U}^{\text{t}}, \mathbf{C}, \mathbf{D}, \mathbf{Q}, \mathbf{P}\right)$;
            \If {$R_{\text{sum}}\geq R_{\text{sum}}^{(\tau)}$}
            \State $\mathbf{Q}^{(\tau+1)}=\mathbf{Q}$, $\mathbf{P}^{(\tau+1)}=\mathbf{P}$, $\mathbf{U}^{\text{r}(\tau+1)}=\mathbf{U}^{\text{r}}$, $\mathbf{U}^{\text{t}(\tau+1)}=\mathbf{U}^{\text{t}}$, $\mathbf{C}^{(\tau+1)}=\mathbf{C}$, $\mathbf{D}^{(\tau+1)}=\mathbf{D}$;
            \Else
            \State $\mathbf{Q}^{(\tau+1)}=\mathbf{Q}^{(\tau)}$, $\mathbf{P}^{(\tau+1)}=\mathbf{P}^{(\tau)}$, $\mathbf{U}^{\text{r}(\tau+1)}=\mathbf{U}^{\text{r}(\tau)}$, $\mathbf{U}^{\text{t}(\tau+1)}=\mathbf{U}^{\text{t}(\tau)}$, $\mathbf{C}^{(\tau+1)}=\mathbf{C}^{(\tau)}$, $\mathbf{D}^{(\tau+1)}=\mathbf{D}^{(\tau)}$;
            \EndIf
            \State Update $\tau=\tau+1$;
        \Until $\frac{|R_{\text{sum}}^{(\tau)}-R_{\text{sum}}^{(\tau-1)}|}{R_{\text{sum}}^{(\tau-1)}}<\delta''$
		\State   Return the optimal solution $\mathbf{Q}^*$, $\left\{\mathbf{U}^{\text{r}*}, \mathbf{U}^{\text{t}*}, \mathbf{C}^*, \mathbf{D}^*\right\}$, $\mathbf{P}^*$.
	\end{algorithmic}  
\end{algorithm}

\subsubsection{Convergence analysis} Given an initial decoding order, the UAV trajectory $\mathbf{Q}$, the active STAR-RIS amplification gain and passive beamforming $\left\{\mathbf{U}^{\text{r}}, \mathbf{U}^{\text{t}}, \mathbf{C}, \mathbf{D} \right\}$, and the power allocation coefficients $\mathbf{P}$ in problem \eqref{eq: optimization_problem} are addressed alternatively, leading to the following inequality:
\begin{equation}
\label{eq: achievable rate convergence analysis}
   \begin{split}
    &R_{\text{sum}}\left(\mathbf{U}^{\text{r}(\tau)},\mathbf{U}^{\text{t}(\tau)}, \mathbf{C}^{(\tau)}, \mathbf{D}^{(\tau)}, \mathbf{Q}^{(\tau)}, \mathbf{P}^{(\tau)}\right)\\
    &\overset{(a)}{\operatorname*{\leq}} R_{\text{sum}}\left(\mathbf{U}^{\text{r}(\tau)},\mathbf{U}^{\text{t}(\tau)}, \mathbf{C}^{(\tau)}, \mathbf{D}^{(\tau)}, \mathbf{Q}^{(\tau+1)}, \mathbf{P}^{(\tau)}\right)\\
    &\overset{(b)}{\operatorname*{\leq}} R_{\text{sum}}\left(\mathbf{U}^{\text{r}(\tau+1)},\mathbf{U}^{\text{t}(\tau+1)}, \mathbf{C}^{(\tau+1)}, \mathbf{D}^{(\tau+1)}, \mathbf{Q}^{(\tau+1)}, \mathbf{P}^{(\tau)}\right)\\
    &\overset{(c)}{\operatorname*{\leq}} R_{\text{sum}}\left(\mathbf{U}^{\text{r}(\tau+1)},\mathbf{U}^{\text{t}(\tau+1)}, \mathbf{C}^{(\tau+1)}, \mathbf{D}^{(\tau+1)}, \mathbf{Q}^{(\tau+1)}, \mathbf{P}^{(\tau+1)}\right).\\
   \end{split}
\end{equation}
Inequality $(\text{a})$ is valid because, with the fixed values of $\left\{\mathbf{U}^{\text{r}(\tau)},\mathbf{U}^{\text{t}(\tau)}, \mathbf{C}^{(\tau)}, \mathbf{D}^{(\tau)}\right\}, \mathbf{P}^{(\tau)}$, the optimal $\mathbf{Q}^{(\tau+1)}$ is found using Algorithm 2; inequality $(\text{b})$ applies as the updated matrices $\left\{\mathbf{U}^{\text{r}(\tau+1)},\mathbf{U}^{\text{t}(\tau+1)}, \mathbf{C}^{(\tau+1)}, \mathbf{D}^{(\tau+1)}\right\}$ are optimally obtained using \textbf{Algorithm 1} with the given matrices of $\mathbf{Q}^{(\tau+1)}$ and $\mathbf{P}^{(\tau)}$; inequality $(\text{c})$ holds since \textbf{Algorithm 3} is employed to determine the optimal power allocation $\mathbf{P}^{(\tau+1)}$ with the provided values of $\left\{\mathbf{U}^{\text{r}(\tau+1)},\mathbf{U}^{\text{t}(\tau+1)}, \mathbf{C}^{(\tau+1)}, \mathbf{D}^{(\tau+1)}\right\}$,and $\mathbf{Q}^{(\tau+1)}$.

The inequalities in \eqref{eq: achievable rate convergence analysis} indicate that the objective function values of \eqref{eq: optimization_problem of Beamforming D}, \eqref{eq:optimization_problem of UAV Trajectory C}, and \eqref{eq:optimization_problem power allocation C} increase monotonically with each iteration of their respective algorithms. Thus, the objective function for \eqref{eq: optimization_problem} does not decrease with each iteration of \textbf{Algorithm 4}. Given that the objective function is constrained by a finite value due to limited transmit power, the convergence of \textbf{Algorithm 4} to a stationary point of \eqref{eq: optimization_problem} is ensured. This condition guarantees that \textbf{Algorithm 4} will reach a stationary point.

\subsubsection{Complexity analysis} The overall complexity of \textbf{Algorithm 4} is primarily determined by the complexities of \textbf{Algorithms 1}, \textbf{2}, and \textbf{3}. For \textbf{Algorithm 1}, the major computational load arises from addressing the relaxed form of problem \eqref{eq: optimization_problem of Beamforming A} within the inner loop. This relaxed problem, being a standard SDP, has a computational complexity of $\mathcal{O}\left(T(K^2+M^{6})\right)$, where $I_{\text{in}}$ and $I_{\text{out}}$ represent the required number of the inner and outer iterations for convergence, respectively. Consequently, the total complexity of \textbf{Algorithms 1} is $\mathcal{O}\left(I_{\text{out}}I_{\text{in}}N(K^2+M^{6})\right)$. Similarly, \textbf{Algorithm 2} and \textbf{Algorithm 3} exhibit a complexity of $\mathcal{O}\left(NK^2+(3N+K)^{3.5}\right)$ and $\mathcal{O}\left(NK^3\right)$. Therefore, the overall complexity of \textbf{Algorithm 4} can be expressed as $\mathcal{O}\left(I_{\text{max}}(I_{\text{out}}I_{\text{in}}N(K^2+M^{6})+(3N+K)^{3.5}+NK^3)\right)$, where $I_{\text{max}}$ indicates the number of iterations required for the convergence of the AO-based iterative algorithm.


\section{Numerical Results}
In this section, we illustrate and discuss the numerical results related to the downlink performance of the UAV-mounted active STAR-RIS system. The positions of the IoT devices are generated randomly within the rectangular region of $800\times500$ meters. For the elevation AoAs/AoDs of \eqref{eq: array response vector}, we set $\omega^{\text{bs}}=\text{arccos} \frac{y_{\text{B}}-y_{\text{u}}}{||\mathbf{q}_{\text{B}}-\mathbf{q}_{\text{u}}||}$ and $\omega^p=\text{arccos} \frac{y_k-y_{\text{u}}}{||\mathbf{q}_k-\mathbf{q}_{\text{u}}||}, p\in \{\text{r}, \text{t}\}$. For the azimuth AoAs/AoDs of \eqref{eq: array response vector}, we define $\psi^{\text{bs}}=\text{arccos} \frac{x_{\text{B}}-x_{\text{u}}}{\sqrt{(x_{\text{B}}-x_{\text{u}})^2+(z_{\text{B}}-H)^2}}$ and $\psi^p=\text{arccos} \frac{x_k-x_{\text{u}}}{\sqrt{(x_k-x_{\text{u}})^2+H^2}}, p\in \{\text{r}, \text{t}\}$. The values of the simulation parameters are listed in details in Table \ref{tab: system parameter} unless otherwise stated.

\begin{table}[h]
\centering
\caption{Simulation Parameters}
\label{tab: system parameter}
\begin{tabular}{ccc}
\hline
Parameter Description                                                  & Value \\ \hline
Number of IoT devices $K$                                    & 6     \\
Location of BS $\mathbf{q}_{\text{B}}$                   &(5,450,5) meters \\  
Initial location of UAV $\mathbf{q}_{\text{u}}^0$    &(0.1,200,30) meters \\  
End location of UAV $\mathbf{q}_{\text{u}}^{\text{F}}$    &(800,200,30) meters \\  
Maximum UAV speed $V_{\text{max}}$        & 11 m/s      \\
BS maximum transmit power $P_{\text{B}}^{\text{max}}$  & 40 dBm      \\
Active STAR-RIS maximum power $P_n^{\text{max}}$    & -20 dBm      \\
Noise power at active STAR-RIS $\sigma_{\text{v}}^2$     &-70 dBm       \\
Noise power at IoT devices $\sigma_{\text{r}}$,$\sigma_{\text{t}}$         & -90 dBm     \\
Pass loss at 1 m $\rho_0$    & -30 dB      \\
Pass loss exponent $\tau_0$    & 2.8    \\
Rician factor $\beta_{\text{bs}}$, $\beta_{\text{s}}$  &  5 dB     \\
Duration of each time slot $\delta$   & 1 s    \\ \hline
\end{tabular}
\end{table}

To demonstrate the superiority of our proposed UAV-aided IoT networks with the active STAR-RIS, we adopt the following five baseline schemes for performance comparison.
\begin{itemize}
    \item \textbf{Baseline scheme 1 (STAR-NOMA):} In this case, the active STAR-RIS is replaced by the conventional passive STAR-RIS operating in the ES mode.  
    \item \textbf{Baseline scheme 2 (RIS-NOMA):} In this case, deploy one conventional reflecting-only RIS and one transmitting-only RIS side by side at the same location to cover the full space. To ensure a fair comparison, each conventional RIS is equipped with $M/2$ elements, assuming $M$ is an even number for simplicity. This configuration can be considered a special case of the passive STAR-RIS, where half of the elements work in the T mode while the other half operate in the R mode.    
    \item \textbf{Baseline scheme 3 (ASTAR-OMA):} In this case, the active STAR-RIS-aided system operates under the OMA protocol. The IoT devices are served with frequency division multiple access (FDMA) to avoid co-carrier interference.
    \item \textbf{Baseline scheme 4 (STAR-OMA):} In this case, the conventional passive STAR-RIS is employed with the FDMA protocol.
    \item \textbf{Baseline scheme 5 (ASTAR-random phase):} In this case, we set all phase shifts randomly, and only optimize the active STAR-RIS beamforming and the UAV trajectory.
\end{itemize}

\begin{figure}[htbp]
\centering
\centerline{\includegraphics[scale=0.65]{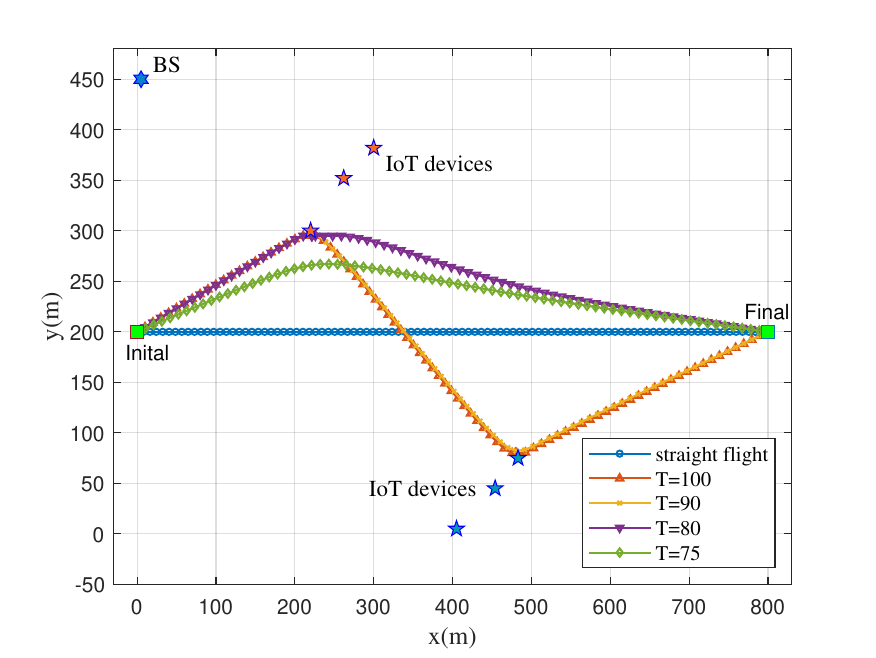}}
\caption{UAV trajectory under different flight times.}
\label{fig:UAV trajectory}
\end{figure}

Fig.~\ref{fig:UAV trajectory} demonstrates the UAV trajectory obtained by the AO algorithm under different flight time, where the number of active STAR-RIS elements is set to $M=10$, and the BS maximum transmit power is configured as $P_{\text{max}}=40$~dBm. It is observed that, as the total flight time increases from $T=75$~s to $T=100$~s, the UAV adjusts its trajectory to enhance communication effectiveness, strategically leveraging the active STAR-RIS's capabilities to optimize coverage and data rates. Specifically, at the beginning of the flight, the UAV maneuvers towards areas with higher IoT device density. This phase of the flight is critical as positioning the UAV nearer to these IoT devices maximizes the efficacy of the reflected/transmitted signal paths, thereby boosting the communication rate. The second part of the flight is to move towards the destination point to meet the flight time constraint. For instance, under shorter flight time like $T=75$~s, the UAV trajectory deviate not much from the direct line between the initial and final points so as to arrive at the destination in time. When longer duration, such as $T=100$~s, is allowed, the UAV is encouraged to move closer to the IoT devices, ensuring that the devices benefit from better signal quality.

\begin{figure}[htbp]
\centering
\centerline{\includegraphics[scale=0.65]{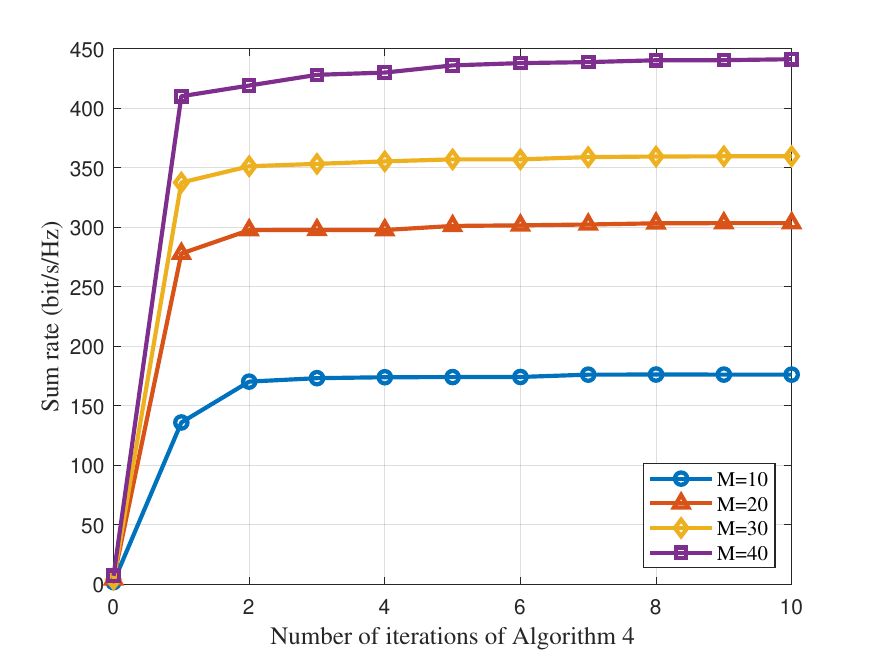}}
\caption{Convergence of the proposed AO-based algorithm for different values of the active STAR-RIS elements.}
\label{fig:convergence}
\end{figure}

Fig.~\ref{fig:convergence} depicts the convergence performance of the proposed AO-based iterative algorithm under different number of the active STAR-RIS elements, i.e, $M$, with the maximum BS transmit power set to $P_{\text{max}}=40$~dBm. These results are obtained from a single random channel realization for each scenario. Observe that the AO algorithm can converge to a stationary state within a small number of iterations. For example, the AO algorithm is stabilized within around $6$ iterations when $M$ is set to $40$. It is also observed that the convergence rate gets slower with the increment of $M$, which is expected as a larger $M$ can introduce higher dimension of the beamforming vectors and thus more complex solving procedure. Moreover, it is also shown that the achievable sum rate increases with the number of $M$. This can be explained by the fact that, the channel gain is enhanced with a larger number of active STAR-RIS elements. 
\begin{figure}[htbp]
\centering
\centerline{\includegraphics[scale=0.65]{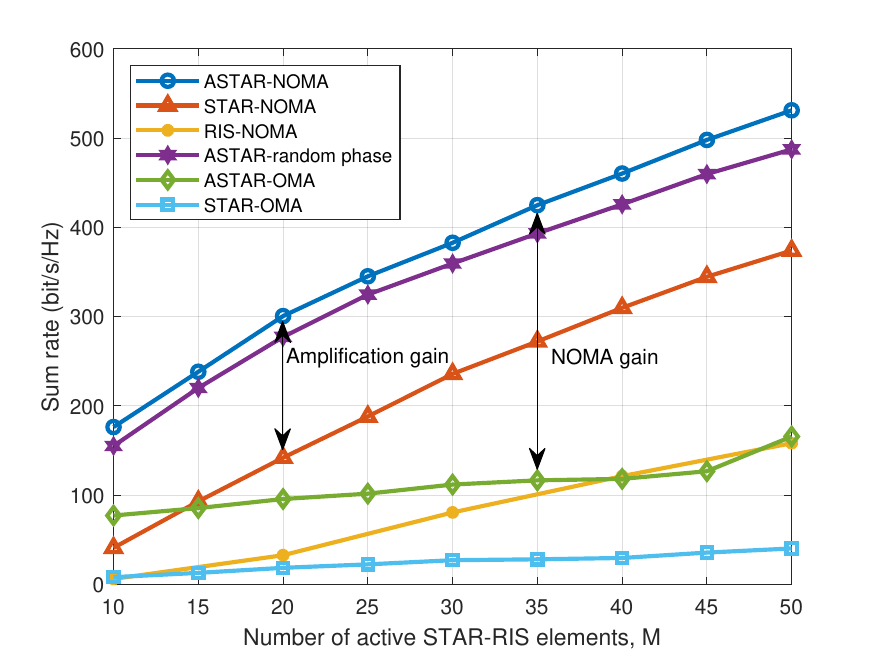}}
\caption{Sum rate versus number of the active STAR-RIS elements.}
\label{fig:M_difference}
\end{figure}

Fig.~\ref{fig:M_difference} portrays the sum rate performance of the AO-based algorithm with the increment of the number of elements i.e. $M$, where the BS maximum transmit power is set to $40$~$\text{dBm}$. One can first observe that the active STAR-RIS NOMA scheme consistently delivers outstanding performance across all $M$ values. This is because, on the one hand, the active amplification provides extra signal enhancement and overcomes the multiplicative path loss brought by the cascaded channel. On the other hand, the NOMA protocol allows the simultaneous transmission over the same resource block, thereby significantly improving the spectrum efficiency. It is also observed that the OMA-based schemes perform significantly worse than their NOMA counterparts, exhibiting lower sum rates and slower growth as $M$ increases. This suggests that the further increment of $M$ would yield limited sum-rate improvement, which underscores the importance of employing NOMA for supporting a large number of IoT devices. The ``RIS-NOMA" scheme shows the lowest sum rate in all the NOMA-based scenarios, indicating that without the specific enhancements with the full-space STAR-RIS, the effectiveness of the UAV-IoT network is significantly diminished. This is due to the reason that the degree-of-freedom (DoF) for adjusting the reflection/transmission configuration is limited with the conventional RIS. Moreover, the ``ASTAR-random phase" scheme shows inferior performance compared to the proposed scheme where the active STAR-RIS phases are optimized, which proves the effectiveness of our proposed AO algorithm.

\begin{figure}[htbp]
\centering
\centerline{\includegraphics[scale=0.65]{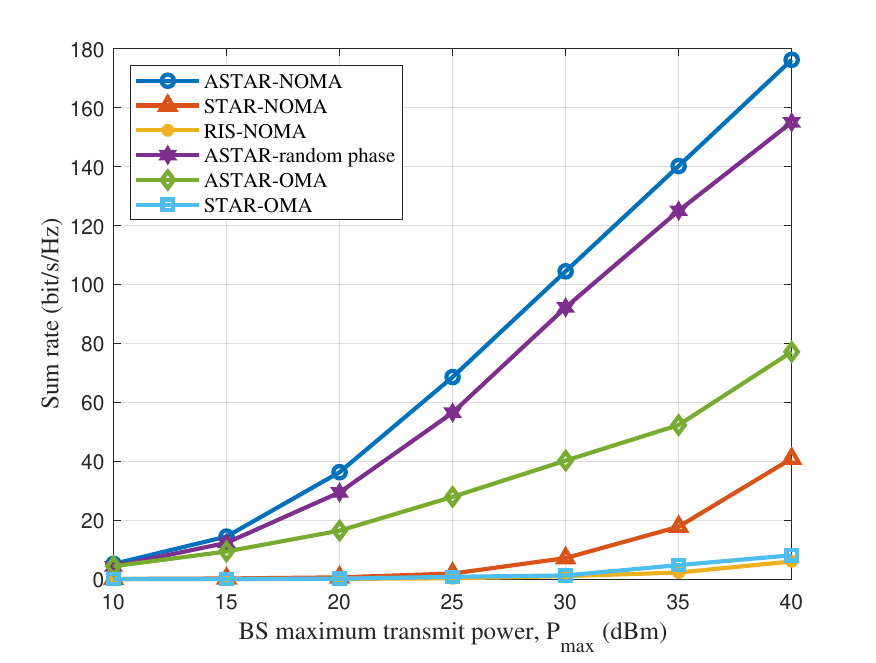}}
\caption{Sum rate versus BS maximum transmit power.}
\label{fig:P_difference}
\end{figure}
In Fig.~\ref{fig:P_difference}, we investigate the sum rate versus the BS maximum transmit power $P_{\text{max}}$, where the number of the active STAR-RIS elements is set to $M=10$. As can be observed, the sum rate of all schemes increases with the increment of $P_{\text{max}}$. Notably, the proposed active STAR-RIS-NOMA scheme exhibits the most significant improvement, and the performance gain of the active STAR-RIS over the passive STRA-RIS becomes more pronounced as the maximum transmit power increases. It can also be observed that the OMA schemes consistently lag behind the NOMA counterparts. The random phase configuration generates performance loss compared to the proposed algorithm.   

\begin{figure}[htbp]
\centering
\centerline{\includegraphics[scale=0.65]{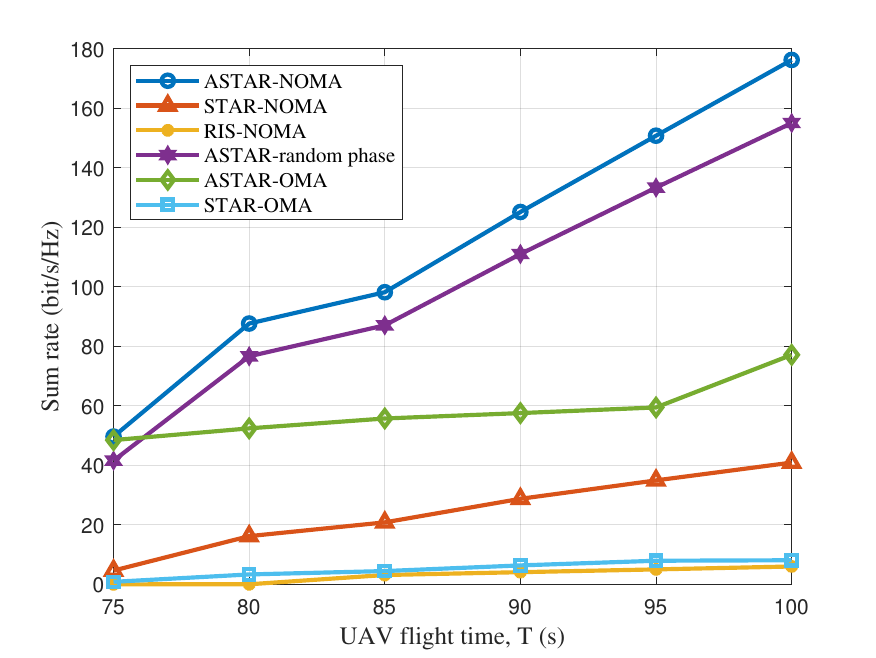}}
\caption{Sum rate versus UAV flight time.}
\label{fig:T}
\end{figure}

Fig.~\ref{fig:T} demonstrates the sum rate versus the UAV flight time $T$, where $P_{\text{B}}^{\text{max}}=40$~dBm and $M=10$. On the one hand, one can observe that the sum rate increases with the flight time. On the other hand, the performance gap between the proposed active STAR-RIS-NOMA scheme and the benchmarks is enlarged with longer flight time, while the performance of conventional RIS-based counterpart shows slight improvement due to the limited DoF. This implies the superiority of deploying the active STAR-RIS in the dynamic scenarios, benefiting by the full-space coverage.

\begin{figure}[htbp]
\centering
\centerline{\includegraphics[scale=0.65]{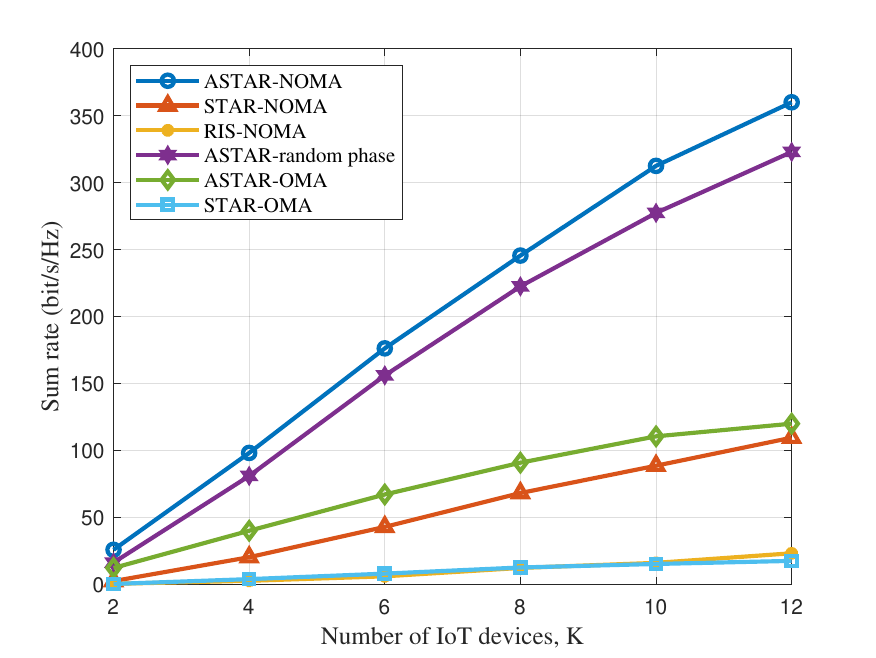}}
\caption{Sum rate versus number of IoT devices.}
\label{fig:user_difference}
\end{figure}

Fig.~\ref{fig:user_difference} illustrates the sum rate versus the number of IoT devices. It is clear to see that, the sum rate of the NOMA schemes consistently increases with the growing number of devices, where the proposed scheme shows the most significant improvement. In contrast, the performance of the OMA schemes is relatively bounded and tends to saturate to a stationary point when the number of IoT devices gets large. This is primarily due to the constraint of the orthogonal resource allocation, which leads to the limited number of connected devices. The obtained results reveal that the NOMA scheme is preferable for supporting the data communication with a large number of IoT devices.

\section{Conclusion}
In this paper, we investigated an IoT communication system with the aid of a UAV-mounted active STAR-RIS and NOMA. With the aim of the sum-rate maximization, the active STAR-RIS beamforming, the UAV trajectory design and the power allocation were jointly optimized. To solve the resulting non-convex problem, an AO-based algorithm was proposed to decouple the original problem into three subproblems. Subsequently, the penalty-based method and the SCA technique were invoked for solving the subproblems to iteratively find the suboptimal solutions. Numerical results demonstrated that the proposed algorithm could significantly improve the sum rate compared to the benchmarks. The results also revealed that the communication links between the BS and the IoT devices could be distinctly enhanced, through leveraging the high-quality channels constructions as well as the extra power compensation. 

However, several interesting research directions remain. Future work could explore 3D UAV trajectory optimization, investigating how to optimize UAV trajectories in three-dimensional space to improve system coverage and performance. Additionally, incorporating energy efficiency as an optimization objective, while considering the power consumption of UAV-mounted STAR-RIS and IoT devices, would be crucial for sustainable system designs. Moreover, addressing the challenges posed by imperfect CSI is an important avenue for future research, as developing robust solutions for imperfect channel state information would improve system performance in dynamic and uncertain environments.

\bibliographystyle{IEEEtran}
\bibliography{mybib}
\end{document}